\documentclass[prd,aps,showpacs,preprintnumbers,amsmath,amssymb]{revtex4-2}
\usepackage[utf8]{inputenc}
\usepackage{graphicx}
\usepackage[breaklinks, colorlinks=true]{hyperref}
\usepackage{bm}
\usepackage{physics}
\usepackage{siunitx}
\usepackage{mathtools}

\newcommand{\ri}{\mathrm{i}}
\newcommand{\re}{\mathrm{e}}
\newcommand{\Lag}{\mathcal{L}}
\newcommand{\bp}{{\bm p}}
\newcommand{\br}{{\bm r}}
\newcommand{\bs}{{\bm s}}
\newcommand{\bK}{{\bm K}}
\newcommand{\btau}{{\bm \tau}}
\newcommand{\brho}{{\bm \rho}}
\newcommand{\bDelta}{{\bm\Delta}}
\newcommand{\calG}{\mathcal{G}}
\newcommand{\calO}{\mathcal{O}}

\begin{document}

\title{Energy-momentum tensor form factors and spin density distribution in the nucleon calculated in a quantized Skyrme model with vector mesons}

\author{Kenji Fukushima}
\email{fuku@nt.phys.s.u-tokyo.ac.jp}
\affiliation{Department of Physics, The University of Tokyo, 
  7-3-1 Hongo, Bunkyo-ku, Tokyo 113-0033, Japan}

\author{Tomoya Uji}
\email{uji@nt.phys.s.u-tokyo.ac.jp}
\affiliation{Department of Physics, The University of Tokyo, 
  7-3-1 Hongo, Bunkyo-ku, Tokyo 113-0033, Japan}

\begin{abstract}
We investigate energy-momentum tensor (EMT) form factors and the spatial spin density distribution in the nucleon within a framework of the quantized Skyrme model with vector mesons.
We construct both the canonical and Belinfante improved EMTs and analyze how pseudogauge uncertainty influences local spin and momentum densities while leaving the global nucleon properties unchanged.
Using the inversion formulas from nucleon matrix elements in the forward limit, we extract the form factors, $A(t)$, $D(t)$, and $J(t)$, in both pseudogauges and the additional antisymmetric form factor, $\mathcal{G}(t)$, associated with the canonical EMT\@.  
We find that the pseudogauge choice leads to sizable differences in the local spin and momentum densities.
In particular, the canonical EMT naturally encodes spin density through the antisymmetric tensor structure, while the Belinfante EMT is sensitive to the total angular momentum only.
Our results illustrate explicitly how different pseudogauges correspond to different spatial interpretations of nucleon spin structure within the same underlying dynamics.
These findings provide a concrete model realization of the pseudogauge ambiguity in QCD-inspired nucleon structure and offer useful intuition for interpreting spatial distributions.
\end{abstract}

\maketitle

%%%%%%%
\section{Introduction}
\label{sec:intro}

Electron–Ion Collider (EIC)~\cite{Accardi:2012qut, AbdulKhalek:2021gbh} will enable precision studies of the three-dimensional structure of the nucleon through hard exclusive processes such as deeply virtual Compton scattering (DVCS)~\cite{Ji:1996nm}. In this framework, generalized parton distributions (GPDs)~\cite{Muller:1994ses, Radyushkin:1996ru, Radyushkin:1997ki} provide access to nucleon tomography, correlating longitudinal momentum fractions with transverse spatial distributions~\cite{Diehl:2002he, Burkardt:2002hr, Burkardt:2005hp}; see Refs.~\cite{Goeke:2001tz, Diehl:2003ny, Belitsky:2005qn} for reviews.  The broad kinematic reach and high luminosity of the EIC will allow systematic measurements of these quantities over wide ranges of Bjorken $x$ and momentum transfer, turning qualitative pictures of nucleon structure into quantitative maps.

A central objective of this program is the determination of the nucleon energy-momentum tensor (EMT) form factors~\cite{Polyakov:2018zvc}.  Through GPD sum rules, the EMT form factors, $A(t)$, $B(t)$, and $D(t)$, can be accessed experimentally~\cite{Ji:1996nm, Ji:1996ek}.  These encode, respectively, the momentum distribution of quarks and gluons, the total angular momentum via Ji’s sum rule~\cite{Ji:1996ek}, and the internal pressure and shear-force distributions that ensure mechanical stability~\cite{Polyakov:1999gs, Polyakov:2002yz}.  The EMT therefore provides a direct link between partonic dynamics at high energies and the static mechanical properties of the nucleon, making its precise theoretical characterization essential for fully exploiting the EIC tomography program.

Among the EMT form factors, the $D$-term form factor is central for nucleon mechanics.   It controls the spatial distributions of pressure and shear forces that can be defined from the spatial components of the EMT in an appropriate frame~\cite{Polyakov:1999gs, Goeke:2001tz, Polyakov:2002yz, Polyakov:2018zvc, Lorce:2018egm}.  These mechanical distributions are not only descriptive but also constrained by stability requirements.   In particular, the pressure profile must satisfy a von~Laue condition ensuring that outward and inward forces compensate~\cite{Polyakov:2002yz, Polyakov:2018zvc, Lorce:2018egm}.   Determining the $D$-term is therefore a direct way to probe the dynamics that confine and bind the constituents into a stable hadron.

Motivated by this connection, the nucleon $D$-term has been the subject of extensive and ongoing work across complementary approaches, including model studies~\cite{Ji:1997gm, Schweitzer:2002nm, Cebulla:2007ei, Goeke:2007fp, Goeke:2007fq, Wakamatsu:2007uc, Kim:2012ts, Jung:2013bya, Hudson:2017oul, Neubelt:2019sou, Chakrabarti:2020kdc, Fujita:2022jus, GarciaMartin-Caro:2023klo, GarciaMartin-Caro:2023toa, Goharipour:2025lep, Sugimoto:2025btn, Tanaka:2025pny, Stegeman:2025sca, Stegeman:2025tdl, Deng:2025azp, Deng:2026gik, Broniowski:2025ctl}, lattice-QCD determinations of EMT form factors and mechanical distributions~\cite{Shanahan:2018nnv, Hackett:2023rif}, phenomenological extractions from DVCS or vector meson photoproduction measurements~\cite{Burkert:2018bqq, Freese:2021qtb, Burkert:2023wzr, Wang:2023fmx}, and data-driven approach~\cite{Cao:2024zlf, Alharazin:2026wfh}.  At the same time, the $D$-term remains less constrained than momentum- and charge-related observables because it is driven by the detailed momentum-transfer dependence of exclusive measurements. The EIC is expected to improve this situation by providing a wider lever arm in both momentum transfer and hard scale, enabling global analyses that reduce model dependence and sharpen the extraction of mechanical information.

The EMT is also a cornerstone of the nucleon spin program.  Moments of GPDs in hard exclusive reactions provide access to the angular momentum carried by quarks and gluons, linking experimentally constrained amplitudes to operator definitions of partonic angular momentum~\cite{Ji:1996ek, Diehl:2003ny, Belitsky:2005qn}.  With polarized beams, the EIC will extend these constraints across flavor and across Bjorken $x$, including sea-quark and small-$x$ regions, and will enable transverse-plane imaging from GPDs that can be viewed as a form of angular-momentum tomography complementary to one-dimensional helicity and unpolarized parton densities~\cite{Lorce:2017wkb, Schweitzer:2019kkd, Liu:2022fvl, Lorce:2025pxt}.

A complete spin decomposition goes further and aims to separate intrinsic spin and orbital motion for quarks and gluons.  Well-known examples include the Ji decomposition~\cite{Ji:1996ek} and the Jaffe-Manohar decomposition~\cite{Jaffe:1989jz}, which differ in how they assign spin and orbital angular momentum, in particular through the distinction between kinetic and canonical OAM.  In a gauge theory, however, such a separation is subtle because the operator definition is not unique.  The question of how to formulate a gauge-invariant angular-momentum decomposition has therefore been discussed for a long time, and it is now understood that gauge-invariant extensions can be constructed through nonlocal operators involving Wilson lines~\cite{Chen:2008ag, Wakamatsu:2010cb, Hatta:2011zs, Hatta:2011ku, Leader:2013jra, Lorce:2015lna}.  In this framework, OAM can be characterized through specific moments of generalized transverse-momentum dependent distributions (GTMDs)~\cite{Meissner:2009ww, Lorce:2011kd}, or in terms of genuine twist-3 generalized parton correlators~\cite{Hatta:2012cs}.  These quantities are expected to be constrained, at least in part, by future measurements at the EIC~\cite{Boussarie:2019icw, Kovchegov:2019rrz, Engelhardt:2020qtg, Bhattacharya:2022vvo}.

These issues are closely connected to the pseudogauge freedom of the EMT~\cite{Speranza:2020ilk}.  Local densities derived from the EMT are not unique because of surface terms that leave conserved charges unchanged but modify the spatial distribution of momentum and angular momentum~\cite{Becattini:2011ev}.  The canonical and Belinfante-Rosenfeld constructions are the best-known examples, but other choices such as the de~Groot-van~Leeuwen-van~Weert formulation could be preferred in the context of spin hydrodynamics~\cite{Florkowski:2018fap}.  Recently, the interpretation of pseudogauge degrees of freedom has invoked intensive discussions for various (seemingly) inequivalent formulations of spin hydrodynamics, e.g., see Refs.~\cite{Fukushima:2020ucl,Li:2020eon,Buzzegoli:2024mra,Drogosz:2024rbd,Fang:2025aig,Becattini:2025twu} for recent developments.   They agree on the total four momentum and total angular momentum, but they differ in how spin and orbital contributions are represented at the density level, and therefore in how one interprets spatial distributions and local decompositions~\cite{Lorce:2015lna}.  This pseudogauge aspect becomes especially important when discussing mechanical properties and pressure distributions, since these rely on specific components of the EMT and on their local interpretation~\cite{Lorce:2015lna, Fukushima:2025jah}.

In the present work, we address these questions in a field-theoretical model of the nucleon based on a quantized Skyrme soliton including explicit vector fields~\cite{Adkins:1983nw}.  Based on collective-coordinate quantization, we compute the nucleon EMT form factors and the associated three-dimensional spatial distributions of angular momentum density, and we present a density-level separation into orbital and spin contributions.  The presence of vector degrees of freedom is particularly useful for this purpose, since it allows us to construct and compare both the canonical and Belinfante versions of the EMT within the same dynamical framework.  It is known that in this model the axial coupling, $g_A$, comes out smaller than the experimental value~\cite{Meissner:1986js}, which limits the scope for fully quantitative phenomenology.  Nevertheless, the framework remains well-suited to our present purpose, namely to explore and visualize the characteristic three-dimensional patterns of the nucleon angular-momentum density.  While EMT form factors in this class of models have been studied previously~\cite{Jung:2013bya}, the role of pseudogauge freedom has not been extensively analyzed yet.  We note, however, that in this model we first minimize the classical soliton mass and only afterwards add the collective rotational energy~\cite{Cebulla:2007ei, Kim:2012ts}.  This treatment does not correspond to an exact solution of the full dynamics, and therefore, the EMT is not exactly conserved.  This is an example of the general problem between variation-after-projection and projection-after-variation, often discussed in nuclear physics.
Since the collective-rotation contribution is subleading in the present work, the lack of exact conservation does not affect our qualitative and semi-quantitative discussions.

Finally, let us add a comment on the pseudogauge freedom.
The canonical and Belinfante tensors are related by pseudogauge transformations, agreeing on the conserved charges, and they coincide in the forward limit.  However, they need not coincide for local densities and at the off-forward form-factor level.  It should be noted that the leading-twist information accessed through twist-2 GPDs effectively projects the ``good'' light-front component, $T^{++}$, for which the Belinfante improvement term vanishes identically~\cite{Lorce:2015lna}.  As a result, the leading-twist observables relevant for the present analysis are insensitive to the pseudogauge choice, even though pseudogauge dependence remains essential for a complete discussion of mechanical properties and for EMT components beyond the $++$ projection.  In practice, sensitivity to such effects requires going beyond leading twist, for example, through higher-twist GPD structures or through transverse-momentum dependent correlators such as $k_T$-moments of TMDs and, more generally, GTMDs~\cite{Lorce:2015lna}.

We work in Minkowski space with the metric convention, $g_{\mu\nu}=\mathrm{diag}(1,-1,-1,-1)$.  We use the symmetrization and anti-symmetrization symbols defined as $a^{(\mu}b^{\nu)}\coloneqq(a^\mu b^\nu + a^\nu b^\mu)/2$ and $a^{[\mu}b^{\nu]}\coloneqq(a^\mu b^\nu - a^\nu b^\mu)/2$, respectively.

%%%%%%%%%%
\section{Energy-momentum tensor and form factors}
\label{sec:general_framework}

This section collects the definitions and kinematic conventions used throughout the paper.  We aim not to review the formalism in full detail, but to clarify the notation to make the paper self-contained.

%%%%%
\subsection{Canonical and Belinfante EMTs and angular-momentum operators}
\label{sec:definitions_emt_spin}

For a generic Lagrangian density, $\Lag(\phi_a,\partial_\mu\phi_a)$, the conserved Noether current associated with translational symmetry reads:
\begin{equation}
  T_{\text{can}}^{\mu\nu}=\sum_a\pdv{\Lag}{(\partial_\mu \phi_a)}\partial^\nu\phi_a-g^{\mu\nu}\Lag\,.
  \label{eq:Tcan_def}
\end{equation}
We define the spin current tensor, $S^{\lambda\mu\nu}$, from the response of $\Lag$ under an infinitesimal Lorentz transformation only on the field $\phi_a$.  In general, $S^{\lambda\mu\nu}$ is not conserved, but iff we apply Noether's formula to this infinitesimal transformation, the canonical spin current reads:
\begin{equation}
  S_{\text{can}}^{\lambda\mu\nu} = \sum_a \pdv{\Lag}{(\partial_\lambda \phi_a)}\,(\Sigma^{\mu\nu}\phi)_a\,,
  \label{eq:spin_current_def}
\end{equation}
where $\Sigma^{\mu\nu}$ denotes the Lorentz generator in the representation of $\phi_a$.  Since $\Sigma^{\mu\nu}$ is anti-symmetric, $S_{\text{can}}^{\lambda\mu\nu} = - S_{\text{can}}^{\lambda\nu\mu}$ holds.
The Lorentz transformation should involve terms from the coordinate transformation, yielding the additional orbital components $L_{\text{can}}^{\lambda\mu\nu}$, and the total angular momentum current, $J_{\text{can}}^{\lambda\mu\nu} = L_{\text{can}}^{\lambda\mu\nu} + S_{\text{can}}^{\lambda\mu\nu}$ that satisfies $\partial_\lambda J_{\text{can}}^{\lambda\mu\nu}=0$.  Interestingly, with Eq.~\eqref{eq:Tcan_def}, the orbital part is represented in terms of the EMT as
\begin{equation}
  L_{\text{can}}^{\lambda\mu\nu} = x^\mu T_{\text{can}}^{\lambda\nu} - x^\nu T_{\text{can}}^{\lambda\mu}\,.
  \label{eq:LxT}
%\label{eq:Jam_LplusS}
\end{equation}
In other words, once the EMT, $T^{\mu\nu}$, and the angular momentum current tensor, $J^{\lambda\mu\nu}$, are given in general, we can identify the orbital part using Eq.~\eqref{eq:LxT}, so that we can extract the spin current tensor, $S^{\lambda\mu\nu}$.  For $S^{\lambda\mu\nu}$ defined in this way, it is easy to show
\begin{equation}
  T^{[\mu\nu]} = \frac{1}{2}\partial_\lambda S^{\lambda\mu\nu}\,,
\label{eq:Tantisym_divS}
\end{equation}
using $\partial_\mu T^{\mu\nu}=0$ and $\partial_\lambda J^{\lambda\mu\nu}=0$.
Equation~\eqref{eq:Tantisym_divS} has a transparent physical interpretation that the antisymmetric part in the EMT is the source to violate the spin current conservation.

It is a well-known problem that the decomposition of $J$ into $L$ and $S$ is not unique, which is rooted in superpotential uncertainty in the EMT definition.  More specifically, let $\Phi^{\lambda\mu\nu}$ be an anti-symmetric tensor with respect to the last two indices, i.e., $\Phi^{\lambda\mu\nu}=-\Phi^{\lambda\nu\mu}$.  From $\Phi^{\lambda\mu\nu}$ another tensor is introduced as
\begin{equation}
  G^{\lambda\mu\nu} = \frac{1}{2}\left(\Phi^{\lambda\mu\nu}+\Phi^{\mu\nu\lambda}+\Phi^{\nu\mu\lambda}\right)\,,
  \label{eq:superpotential_G}
\end{equation}
which satisfies $G^{\lambda\mu\nu}=-G^{\mu\lambda\nu}$, and thus $\partial_\mu\partial_\lambda G^{\lambda\mu\nu}=0$.
It is then straightforward to confirm that the ``improved'' EMT and angular momentum constructed as
\begin{align}
  T_{\text{imp}}^{\mu\nu} = T^{\mu\nu} + \partial_\lambda G^{\lambda\mu\nu}\,,
    \qquad
  J_{\text{imp}}^{\lambda\mu\nu} = J^{\lambda\mu\nu} + \partial_\rho \left( x^\mu G^{\rho\lambda\nu} - x^\nu G^{\rho\lambda\mu} \right)\,,
  \label{eq:PGtrans}
\end{align}
are qualified as the conserved currents satisfying $\partial_\mu T_{\text{imp}}^{\mu\nu}=0$ and $\partial_\lambda J_{\text{imp}}^{\lambda\mu\nu}=0$.  These relations to redefine the EMT and the angular momentum are referred to as the pseudogauge transformation.  From these expressions, in the same way as above, it is possible to extract the spin part as
\begin{align}
  S_{\text{imp}}^{\lambda\mu\nu} = S^{\lambda\mu\nu} - \Phi^{\lambda\mu\nu}\,.
  \label{eq:pseudogauge_S_Phi}
\end{align}
We note that the pseudogauge transformation in Eq.~\eqref{eq:PGtrans} adds only surface terms, so that the conserved charges integrated over volume are intact as long as the density vanishes sufficiently fast at spatial infinity.  Nevertheless, it is clear that the pseudogauge transformation shifts the spin current and reorganize decomposition of the angular momentum.

Although $\Phi^{\lambda\mu\nu}$ is arbitrarily chosen, it is often the most convenient to make a special choice of $\Phi^{\lambda\mu\nu}=S^{\lambda\mu\nu}_{\text{can}}$, which leads to the well-known Belinfante improved form of the EMT and the angular momentum.  In this case, the EMT turns out to be symmetric, i.e.,
%\begin{equation}
%    T^{\mu\nu}_{\text{Bel}}=T^{\mu\nu}_{\text{can}}+\frac{1}{2}\partial_\lambda\left(S^{\lambda\mu\nu}_{\text{can}}+S^{\mu\nu\lambda}_{\text{can}}+S^{\nu\mu\lambda}_{\text{can}}\right)\,,
%    \qquad
%    T^{\mu\nu}_{\text{Bel}}=T^{\nu\mu}_{\text{Bel}}\,.
%\label{eq:Tbel_def}
%\end{equation}
\begin{align}
  T_{\text{Bel}}^{\mu\nu} = T_{\text{Bel}}^{\nu\mu}\,.
\end{align}
This can be understood from Eqs.~\eqref{eq:Tantisym_divS} and \eqref{eq:pseudogauge_S_Phi}.  Obviously, $S_{\text{Bel}}^{\lambda\mu\nu}=0$, and the spin conservation is trivially satisfied, and therefore $T_{\text{Bel}}^{[\mu\nu]}=0$ is concluded.  The spin current is absorbed into a superpotential and does not appear explicitly in the Belinfante form, leading to the total angular momentum current written solely in terms of the symmetric EMT, that is,
\begin{equation}
  J^{\lambda\mu\nu}_{\text{Bel}} = x^\mu T^{\lambda\nu}_{\text{Bel}}-x^\nu T^{\lambda\mu}_{\text{Bel}}\,.
  \label{eq:Jam_Bel_orbital}
\end{equation}
We will refer to $T^{\mu\nu}_{\text{can}}$ as the canonical EMT and to the symmetric EMT obtained by the Belinfante improvement as the Belinfante EMT in what follows below.

The Belinfante EMT is the standard choice in relativistic field theory.
In gauge theories without fermions, it coincides with the Hilbert EMT obtained from metric variation, i.e.,
\begin{align}
    T_{\text{Hil}}^{\mu\nu} = -\frac{2}{\sqrt{|g|}} \frac{\delta S}{\delta g_{\mu\nu}} \,,
\end{align}
where $S$ is the action.

%%%%%
\subsection{EMT form factors for a spin-$1/2$ target}
\label{sec:emt_ff_general}

We consider EMT matrix elements in the one-nucleon state basis spanned by $\ket{p,s}$.  In this notation, $s$ is the nucleon spin and $p$ represents the four momentum, $p^\mu=(E, \bp)$, where $E=\sqrt{|\bp|^2+M^2}$ with $M$ the nucleon mass.  We adopt the normalization convention as
\begin{equation}
    \braket{p^\prime,s^\prime}{p,s} = 2E_p(2\pi)^3\delta^{(3)}(\bm{p}^\prime-\bm{p})\delta_{s^\prime s}\,.
\label{eq:state_norm}
\end{equation}
Since we discuss not only the symmetric Belinfante EMT but also the non-symmetric canonical EMT, we need to introduce the structure functions corresponding to the decomposition of $T^{\mu\nu}=T^{(\mu\nu)}+T^{[\mu\nu]}$.

For the symmetric part of the EMT, the most general decomposition of the nucleon matrix element consistent with invariance under Lorentz, parity, and time-reversal transformations takes the following form:
\begin{align}
    \mel{p^\prime, s^\prime}{T^{(\mu\nu)}(x)}{p, s} \eqqcolon \bar{u}(p^\prime,s^\prime) \Biggl[ A(t)\gamma^{(\mu}P^{\nu)} + B(t)\frac{P^{(\mu} \ri \sigma^{\nu)\rho}\Delta_\rho}{2M} + D(t)\frac{\Delta^\mu\Delta^\nu-g^{\mu\nu}\Delta^2}{4M} \Biggr] u(p,s) \, \re^{\ri(p^\prime-p) x} \,,
\label{eq:emt_ff_decomp_sym}
\end{align}
where $P=(p^\prime+p)/2$, $\Delta=p^\prime-p$, and $t=\Delta^2$.
As usual, the spin tensor is defined by $\sigma^{\mu\nu} = \tfrac{\ri}{2}[\gamma^\mu,\gamma^\nu]$ and the spinors are normalized as $\bar{u}(p, s)u(p, s)=2M$.  Generic non-symmetric EMTs can accommodate additional Lorentz structures associated with the anti-symmetric part, $T^{[\mu\nu]}$~\cite{Lorce:2015lna, Won:2025dgc}.
In the present case where $T^{[\mu\nu]}$ is local, parity-even, and time-reversal-even, one extra form factor is sufficient to parameterize $T^{[\mu\nu]}$ as
\begin{align}
    \mel{p^\prime, s^\prime} {T^{[\mu\nu]}(x)}{p,s}
    =\frac{\ri}{2}\Delta_\lambda\mel{p^\prime, s^\prime}{S^{\lambda\mu\nu}(x)}{p, s}
    \eqqcolon \bar u(p^\prime, s^\prime) \Biggl[ \calG(t) \frac{P^{[\mu} \ri\sigma^{\nu]\rho}\Delta_\rho}{2M} \Biggr] u(p, s) \, \re^{\ri(p^\prime-p)x} \,.
\label{eq:emt_ff_decomp_antisym}
\end{align}
%so that the full matrix element reads
%\begin{equation}
%    \mel{p^\prime, s^\prime}{T^{\mu\nu}(x)}{p, s}=\mel{p^\prime, s^\prime}{T^{(\mu\nu)}(x)}{p, s}+\mel{p^\prime, s^\prime}{T^{[\mu\nu]}(x)}{p, s}\,.
%\label{eq:emt_ff_full_decomp}
%\end{equation}
For the relation between the first and second expressions, see Eq.~\eqref{eq:Tantisym_divS}~\cite{Lorce:2015lna}.
We note that $\calG(t)$ identically vanishes for the Belinfante EMT by construction.  As discussed later, the canonical EMT obtained from our Skyrme-model calculation has a property of $\expval{T^{0i}}\neq \expval{T^{i0}}$, while the spatial parts are symmetric, i.e., $\expval{T^{ij}}=\expval{T^{ji}}$.
%Accordingly, we extract $A(t)$, $B(t)$, and $D(t)$ from the symmetric part via Eq.~\eqref{eq:emt_ff_decomp_sym}, and quantify the residual antisymmetric contribution through $\mathcal{G}(t)$ in Eq.~\eqref{eq:emt_ff_decomp_antisym}.

The form factors, $A(t)$, $B(t)$, and $D(t)$ encode the mechanical and angular momentum properties of the target~\cite{Polyakov:2002yz, Polyakov:2018zvc, Lorce:2018egm}.  It is well-known that the momentum conservation imposes the normalization condition, $A(0)=1$.
The combination of $A(t)$ and $B(t)$ defines the angular-momentum form factor~\cite{Ji:1996ek},
\begin{equation}
    J(t)=\frac{1}{2}\bigl[A(t)+B(t)\bigr]\,.
    \label{eq:Jt_def_from_AB}
\end{equation}
If the EMT is symmetric as in the Belinfante-improved form, the normalization of the Poincar\'e generators imposes $J(0)=1/2$ for a spin-$1/2$ target.
%In this symmetric case, the mixed components, $T^{0i}=T^{i0}$, encode the OAM density in the Breit frame.
It should be noted that $D(t)$ encodes the internal mechanical properties of the target, representing the pressure and shear-force distributions~\cite{Polyakov:2018zvc}.

For the non-symmetric EMT as in the canonical form, in contrast, another form factor, $\calG(t)$, appears from the anti-symmetric part, $T^{[\mu\nu]}$.  It should be noted that $J(0)$ is no longer interpreted as the total angular momentum. The OAM density, i.e., $x^iT^{0j}-x^jT^{0i}$, depends on both the symmetric and anti-symmetric parts of the EMT\@.  Although it is known that the OAM charge is given by $J(0)+\tfrac{1}{2}\calG(0)$, we note that $J(t)+\frac{1}{2}\calG(t)$ is not necessarily interpreted as the OAM form factor, for $\calG(t)$ is not a simple function of $S^{0\mu\nu}(x)$ as seen in Eq.~\eqref{eq:emt_ff_decomp_antisym}.  To recover the total angular momentum, $1/2$, the spin-current contribution should be estimated separately.  

%%%%%
\subsection{Relation to partonic observables}
\label{sec:partonic_relation}

We shall remark on the experimental determination of $A(t)$, $B(t)$, and $D(t)$.
The EMT form factors are, in principle, observables, admitting a partonic interpretation in terms of quark and gluon correlation functions.  We discussed two types of definitions of the QCD EMT, i.e., $T^{\mu\nu}_\text{Bel}$ and $T^{\mu\nu}_\text{can}$, and the corresponding form factors may be different in general.  In contrast to gauge-invariant $T^{\mu\nu}_\text{Bel}$, a na\"{i}ve local form of $T^{\mu\nu}_\text{can}$ is gauge variant, and this seems to obstruct its direct connection to measurable observables.  However, one can still extend the canonical EMT to a light-front gauge-invariant form, $T^{\mu\nu}_\text{gic}$, by allowing for a controlled nonlocality~\cite{Chen:2008ag, Hatta:2011zs}.  In this way, both $T^{\mu\nu}_\text{Bel}$ and $T^{\mu\nu}_\text{gic}$ admit operator definitions that can be related to experimentally accessible parton correlation functions, such as GPDs and TMDs; see Ref.~\cite{Lorce:2015lna}.

The relevant partonic information at the leading twist order is contained in the twist-2 unpolarized GPDs, i.e., $H^a(x,\xi,t)$ and $E^a(x,\xi,t)$ with the index $a$ labeling the parton species.  These GPDs can be extracted from the DVCS measurement.  Their second Mellin moments satisfy the well-known $x$-moment sum rules:
\begin{align}
    \label{eq:sum_rule_H}
    \int_{-1}^{1} \dd{x}\, x\,H^a(x,\xi,t) &= A^a(t) + \xi^2 D^a(t)\,, \\
    \int_{-1}^{1} \dd{x}\, x\,E^a(x,\xi,t) &= B^a(t) - \xi^2 D^a(t)\,,
    \label{eq:sum_rule_E}
\end{align}
which directly relate GPDs to the EMT matrix elements.  As seen from the above expressions, the $\calO(\xi^0)$ and $\calO(\xi^2)$ terms determine the EMT form factors of $A^a(t)$ and $B^a(t)$, and the mechanical form factor, $D^a(t)$, respectively.  Technically, these relations follow from matching the second Mellin moments of the twist-2 correlator with a light-front projection along a lightlike direction to the corresponding EMT matrix elements, i.e., $T^{++}=T^{\mu\nu}n_\mu n_\nu$ with $n_\mu\propto (1,0,0,1)$.

At this point, one may wonder which EMT form factors can be determined by the GPD sum rules in Eqs.~\eqref{eq:sum_rule_H} and \eqref{eq:sum_rule_E}.  While the unpolarized GPDs, namely $H^a$ and $E^a$, are defined unambiguously as gauge-invariant parton correlation functions, the EMT form factors appearing on the right-hand side of Eqs.~\eqref{eq:sum_rule_H} and \eqref{eq:sum_rule_E} could be interpreted, \textit{a priori}, as quantities referring to either the Belinfante or canonical EMT\@.  The key point is that the leading-twist relations are insensitive to the difference between $T^{\mu\nu}_\text{Bel}$ and $T^{\mu\nu}_\text{gic}$: twist~2 GPDs probe only the symmetric part of the EMT, and the pseudogauge ambiguity is a total-derivative term that drops out upon projection onto the $++$ components.  Consequently, for any $t$, the twist-2 sum rules determine a common set of leading-twist EMT form factors.  Accessing genuine differences between $T^{\mu\nu}_\text{Bel}$ and $T^{\mu\nu}_\text{gic}$ requires going beyond leading twist, for instance, higher-twist GPDs or transverse-momentum dependent correlators such as TMDs or GTMDs, where the process-dependent Wilson-line geometry becomes essential~\cite{Lorce:2015lna}.

%%%%%%%%%%
\section{Spatial structures and inversion formulas}

We shall explain some useful relations needed for the computation of the EMT form factors and the angular-momentum density in the model construction.

%%%%%
\subsection{Breit-frame densities and spatial structures}
\label{sec:breit_densities}

To connect the EMT form factors to the spatial distributions, it is convenient to work in the Breit frame defined as a frame with $\Delta^0=0$ (and thus $t=-\bm{\Delta}^2$).  We should stress that the interpretation of Breit-frame form factors as three-dimensional spatial densities is subject to well-known limitations~\cite{Lorce:2020onh, Jaffe:2020ebz, Panteleeva:2021iip, Epelbaum:2022fjc}.  In a fully relativistic system, recoil effects prevent a model-independent three-dimensional density interpretation, and transverse light-front densities provide a more robust representation.  In the present work, however, the calculation is performed in a semiclassical large-$N_c$ soliton framework.  The nucleon is treated as a heavy static object, and the densities are computed in the soliton rest frame.  In the heavy-mass limit, recoil effects are suppressed, and the rest-frame densities can be identified with Breit-frame densities as a low-energy approximation.

The static EMT densities or the Breit-frame densities are introduced as results of the three-dimensional Fourier transform, i.e.,
\begin{equation}
    T^{\mu\nu}(\br; \bs, \bs^\prime) = \int\frac{\dd[3]{\bm{\Delta}}}{(2\pi)^3} \, \frac{\re^{-\ri\bDelta\cdot\br}}{2E} \mel{p^\prime, s^\prime}{T^{\mu\nu}(0)}{p,s}\Bigr|_{\Delta^0=0} \,.
\label{eq:breit_density_def}
\end{equation}
In this frame, the nucleon energy is written as $E=\sqrt{M^2+\bm{\Delta}^2/4}$.  In terms of the EMT form factors, the tensor elements of the Breit-frame densities take the following expressions in terms of the form factors:
\begin{align}
    T^{00}(\br; \bs, \bs^\prime) &= M\int\frac{\dd[3]{\bDelta}}{(2\pi)^3} \, \re^{-\ri\bDelta\cdot\br} \biggl\{ A(t)-\frac{t}{4M^2} \bigl[ A(t) - 2J(t)+D(t)\bigr] \biggr\} \delta_{\bs, \bs^\prime}\,,
    \label{eq:breit_density_00} \\
    T^{ij}(\br; \bs, \bs^\prime) &= \frac{1}{4M}\int\frac{\dd[3]{\bDelta}}{(2\pi)^3} \, \re^{-\ri\bDelta\cdot\br} \bigl(\Delta^i\Delta^j-\delta^{ij}\bDelta^2\bigr) D(t) \, \delta_{\bs, \bs^\prime}\,,
    \label{eq:breit_density_ij} \\
    T^{0i}(\br; \bs, \bs^\prime) &= \frac{1}{2}\int\frac{\dd[3]{\bDelta}}{(2\pi)^3} \, \re^{-\ri\bDelta\cdot\br} \bigl(-\ri\bDelta\times\bm{\sigma}_{s^\prime, s}\bigr)^i \Bigl[ J(t)+\frac{1}{2}\calG(t) \Bigr] \,,
    \label{eq:breit_density_0i} \\
    T^{i0}(\br; \bs, \bs^\prime) &= \frac{1}{2}\int\frac{\dd[3]{\bDelta}}{(2\pi)^3} \, \re^{-\ri\bDelta\cdot\br} \bigl(-\ri\bDelta\times\bm{\sigma}_{s^\prime, s}\bigr)^i \Bigl[ J(t)-\frac{1}{2}\calG(t) \Bigr] \,.
    \label{eq:breit_density_i0}
\end{align}
Here, we introduce a notation, $\sigma^i_{s^\prime, s} = \chi^\dagger_{s^\prime}\sigma^i\chi_s$ using the nucleon Pauli spinors, $\chi_{s^\prime}$ and $\chi_s$, normalized as $\chi^\dagger_{s^\prime}\chi_s=\delta_{s, s^\prime}$.  In what follows, we set $\bm s=\bm{s}^\prime=\hat{\bm{z}}$ and $|\bm s|=s=+1$.

From the symmetry, these quantities are further decomposed into five functions of the radial distance, $r=|\br|$.  For spherically symmetric static configurations, we immediately identify:
\begin{align}
    T^{00}(\br) = \varepsilon(r)\,,
\end{align}
where $\varepsilon(r)$ denotes the energy density.  The spatial EMT components or the stress tensor components admit the standard decomposition:
\begin{equation}
    T^{ij}(\bm{r})=\delta^{ij}p(r)+\left(\hat{r}^i\hat{r}^j-\frac{1}{3}\delta^{ij}\right)s(r)\,,
\label{eq:stress_decomp}
\end{equation}
where $\hat{r}^i=r^i/r$.  This decomposition provides an intuitive interpretation of $p(r)$ as the pressure and $s(r)$ as the shear force.  It should be emphasized, however, that the local quantities $p(r)$ and $s(r)$ defined in this way are pseudogauge dependent~\cite{Fukushima:2025jah}.  By contrast, a force-density distribution, known as the Cauchy momentum equation~\cite{Polyakov:2018exb, Won:2023cyd, Freese:2024rkr, Ji:2025gsq, Ji:2025qax, Kim:2025iis}, 
\begin{equation}
    \partial_i T^{ij}(\bm{r}) = \mathcal{F}_j(r)\,,
\label{eq:force_density}
\end{equation}
is free from the pseudogauge ambiguity because pseudogauge transformations modify the EMT by surface terms only.

For a spin-$1/2$ target with the spin quantized along the $z$ axis, i.e., $\bm{s}=s\hat{\bm z}$, parity and time-reversal invariance requires the linear dependence of $T^{0i}$ and $T^{i0}$ on spin, that is,
\begin{equation}
    T^{0i}(\br) = -(\hat{\br} \times \bs)^i \rho_0(r)\,,
    \qquad
    T^{i0}(\br) = -(\hat{\br} \times \bs)^i \rho_1(r)\,.
\label{eq:mixed_rho01}
\end{equation}
It should be noted that $\rho_0(r)\neq \rho_1(r)$ in general for the non-symmetric EMT case.  It is then convenient to make the symmetric and anti-symmetric combinations as
\begin{equation}
    T^{(0i)}(\br) = -(\hat{\br} \times \bs)^i \rho_+(r)\,,
    \qquad
    T^{[0i]}(\br) = -(\hat{\br} \times \bs)^i \rho_-(r)\,,
%    T^{(0i)}(\bm r) = \epsilon^{ijk} s^{j} \hat r^{k} \rho_+(r)\,,
%    \qquad
%    T^{[0i]}(\bm r) = \epsilon^{ijk} s^{j} \hat r^{k} \rho_-(r)
\label{eq:mixed_rhopm}
\end{equation}
using $\rho_\pm(r) = \tfrac{1}{2}[\rho_0(r)\pm \rho_1(r)]$.
%\begin{equation}
%    \rho_\pm(r)=\frac{1}{2}\bigl[\rho_0(r)\pm \rho_1(r)\bigr]\,.
%\label{eq:rhopm_def}
%\end{equation}

%%%%%
\subsection{Inversion formulas: from Breit-frame densities to EMT form factors}
\label{sec:inversion}

In the model studies, as we discuss in the present paper, we first calculate the Breit-frame densities, from which we wish to extract information about the form factors.  Therefore, we need to establish the inversion formulas from the Breit-frame densities to the EMT form factors by means of the Fourier-Bessel transformation.

Let us introduce the Fourier transforms of the static densities:
\begin{equation}
  \widetilde T^{\mu\nu}(\bm\Delta) = \int \dd[3]{r} \re^{\ri\bDelta\cdot\br} \,  T^{\mu\nu}(\br) \,.
  \label{eq:Ttilde_def}
\end{equation}
Plugging Eq.~\eqref{eq:breit_density_def} into Eq.~\eqref{eq:Ttilde_def} leads to
\begin{equation}
  \widetilde T^{\mu\nu}(\bm\Delta) = \frac{1}{2E} \mel{p^\prime, s}{T^{\mu\nu}(0)}{p, s} \Bigr|_{\Delta^0=0} \,.
  \label{eq:Ttilde_me_relation}
\end{equation}
The comparison with Eq.~\eqref{eq:breit_density_ij} gives its relation to the mechanical form factor as
\begin{equation}
%    D(t) = \frac{6M}{t^2}\left(\Delta_i\Delta_j-\frac{1}{3}\delta_{ij}\bm{\Delta}^2\right)\widetilde T^{ij}(\bm\Delta) \,.
    \qquad
    D(t) = \frac{2M}{t} \delta_{ij} \widetilde T^{ij}({\bm\Delta}) \,.
\label{eq:D_from_Tij_proj}
\end{equation}
%For the static case, it is easy to prove $\Delta_i \widetilde{T}^{ij}({\bm \Delta}) = \widetilde{T}^{ij}({\bm \Delta}) \Delta_j = 0$.  In the above expression, we inserted an irrelevant term, $\Delta_i \Delta_j \tilde{T}^{ij}({\bm \Delta})$, to project the traceless part.
%In the right-hand side, $\Delta_i$ should be contracted in the end, so that the whole quantity is a function of $t=-{\bm \Delta}^2$.
After taking the contract of spatial indices, the right-hand side turns out to be a function of $t$ alone.
This feature becomes more evident from the expressions in terms $p(r)$, that is,
%and $s(r)$.  Using the symmetric stress tensor~\eqref{eq:stress_decomp}, if we drop the irrelevant term of $\Delta_i \Delta_j \widetilde{T}^{ij}({\bm \Delta})$, only the term with $p(r)$ is selected.  After the angular integration, we find:
\begin{equation}
    D(t) = \frac{6M}{t} 4\pi \int_0^\infty\dd{r} r^2 \,j_0(\sqrt{-t}r)\,p(r)\,,
    \label{eq:D_from_p}
\end{equation}
where $\sqrt{-t}=|{\bm \Delta}|$ and $j_n(x)$ is the spherical Bessel function.
%Alternatively, we can use Eq.~\eqref{eq:D_from_Tij_proj} to drop the terms $\propto \delta^{ij}$ including the $p(r)$ term, resulting in
Interestingly, it is also possible to express $D(t)$ in terms of $s(r)$ as
\begin{equation}
    D(t) = \frac{4M}{t} 4\pi\int_0^\infty\dd{r} \, r^2 \,j_2(\sqrt{-t}r)\,s(r)\,.
    \label{eq:D_from_s}
\end{equation}
%We note that the order of $j_n(x)$ changes due to angular dependence of $\hat{r}^i \hat{r}^j$.

In a static configuration with $\bm s=\hat{\bm z}$, Eq.~\eqref{eq:mixed_rho01}  implies that the mixed components are fully encoded in $\rho_\pm(r)$.  Indeed, Eqs.~\eqref{eq:breit_density_0i} and \eqref{eq:breit_density_i0} are translated into the following general structures:
\begin{equation}
    \widetilde T^{(0i)}(\bm\Delta) = -\frac{\ri}{2}(\bDelta \times \bs)^i J(t)\,,
    \qquad
    \widetilde T^{[0i]}(\bm\Delta) =  -\frac{\ri}{4}(\bDelta \times \bs)^i \calG(t)\,.
\label{eq:mixed_Ttilde_structures}
\end{equation}
These relations can be inverted to the one-dimensional integration formulas as~\cite{Schweitzer:2019kkd},
\begin{equation}
    J(t) = \frac{8\pi}{\sqrt{-t}}\int_0^\infty \dd{r}r^2j_1(\sqrt{-t}r) \, \rho_+(r)\,,
    \qquad
    \calG(t) = \frac{16\pi}{\sqrt{-t}}\int_0^\infty \dd{r}r^2 j_1(\sqrt{-t}r) \, \rho_-(r)\,.
\label{eq:J_G_from_rhopm}
\end{equation}
% where $j_1$ is a spherical Bessel function.

Defining the Fourier transform of the energy density as
\begin{equation}
    \tilde{\varepsilon}(t) = \widetilde T^{00}(\bDelta)
    = \int \dd[3]{r}  \re^{\ri\bDelta\cdot\br} \, \varepsilon(\br)
    = 4\pi\int_0^\infty \dd{r}\, r^2\, j_0(\sqrt{-t}r) \, \varepsilon(r)\,,
\label{eq:Ecal_def}
\end{equation}
and using Eq~\eqref{eq:breit_density_00}, $\tilde{\varepsilon}$ can be related to other form factors as
%    \mathcal{E}(t)=\frac{1}{M}\widetilde T^{00}(\bm\Delta)
%    =\frac{1}{M}\int \dd[3]{r}\re^{\ri\bm\Delta\cdot\bm r}\varepsilon(\bm r)
%    =\frac{4\pi}{M}\int_0^\infty \dd{r}r^2j_0(\sqrt{-t}r)\varepsilon(r)\,,
%the relation between $\mathcal{E}(t)$ and the form factors reads
\begin{equation}
    \frac{\tilde\varepsilon(t)}{M} = A(t)-\frac{t}{4M^2}\Bigl[A(t)-2J(t)+D(t)\Bigr]\,,
\label{eq:Ecal_relation}
\end{equation}
from which $A(t)$ follows as
\begin{equation}
    A(t) = \frac{M\tilde{\varepsilon}(t) - (t/2) J(t) + (t/4)D(t)}{M^2 - t/4}\,.
\label{eq:A_from_EJD}
\end{equation}
Finally, the remaining symmetric form factor, $B(t)$, is obtained immediately from $J(t)=\tfrac{1}{2}[A(t)+B(t)]$.
%\begin{equation}
%    B(t)=2J(t)-A(t)\,.
%\label{eq:B_from_AJ}
%\end{equation}

%%%%%%%%%
\section{Model setup}

We make an overview of the Skyrme model with vector mesons and the semiclassical framework used in this work.
We first specify the effective Lagrangian and determine a static soliton solution with the baryon number fixed to be $B=1$.  Then, we introduce adiabatic collective rotation, including the rotationally induced vector-meson components, and estimate the moment of inertia.  Finally, we quantize the rotational zero modes and evaluate nucleon matrix elements of the EMT and the spin-current tensor.  For more details about the formulation of the model setup and the classical soliton solution, see Ref.~\cite{Meissner:1986js}.

%%%%%
\subsection{Effective Lagrangian}
\label{sec:model}

We consider the two-flavor chiral field, $U = \xi^2\in SU(2)$, in the nonlinear representation in terms of the pions, coupled to ${\bm\rho}_\mu$ (isotriplet vector mesons) and $\omega_\mu$ (isosinglet vector meson).
The Lagrangian density, $\Lag_{\text{model}} = \Lag_{\pi} + \Lag_{1} + \Lag_{V} + \Lag_{\text{WZW}} + \Lag_{m}$, consists of
\begin{align}
\label{eq:Lagrangian_detail}
    &\Lag_{\pi} =\frac{1}{4}f_\pi^2\tr(\partial_\mu U\partial^\mu U^\dagger)\,, \\
    &\Lag_{1} = -\frac{1}{2}f_\pi^2\tr\Bigl[\bigl(D_\mu\xi\cdot\xi^\dagger + D_\mu\xi^\dagger\cdot\xi\bigr)^2\Bigr]\,,
    \label{eq:L1}\\
    &\Lag_{V} = -\frac{1}{2g^2}\tr(F_{\mu\nu}F^{\mu\nu})\,, \\
    &\Lag_{\text{WZW}} = \frac{3}{2}g \, \omega_\mu B^\mu\,,
    \label{eq:LWZW}\\
    &\Lag_{m} = \frac{1}{4}m_\pi^2 f_\pi^2\tr(U + U^\dagger - 2)\,.
\end{align}
The explicit form of the topological baryon current reads:
\begin{equation}
  B^\mu = \frac{1}{24\pi^2}
  \varepsilon^{\mu\nu\alpha\beta}
  \tr\bigl(U^\dagger\partial_\nu UU^\dagger\partial_\alpha UU^\dagger\partial_\beta U\bigr)\,.
\end{equation}
The covariant derivative and the non-Abelian field strength tensors are defined by
\begin{align}
  D_\mu = \partial_\mu - \ri V_\mu\,,
  \qquad
  F_{\mu\nu} = \partial_\mu V_\nu - \partial_\nu V_\mu - \ri[V_\mu, V_\nu]\,,
\end{align}
where the vector-meson matrix with ${\bm\tau}$ the Pauli matrix in flavor space is given by
\begin{equation}
  V_\mu = \frac{g}{2}\bigl(\bm{\tau}\cdot\bm{\rho}_\mu + \omega_\mu\bigr)\,,
\end{equation}
where $\btau\cdot\brho_\mu = \tau^a \rho^a_\mu$ with the index $a$ in flavor space and the Lorentz index $\mu$.

%%%%%
\subsection{Classical soliton profiles}
\label{sec:classical_profiles}

We shall work in the baryon rest frame and specifically focus on static configurations.  For the pion field, we adopt the hedgehog Ansatz, i.e., $U(\br) = \xi^2(\br)$ with
\begin{equation}
  \label{eq:ansatz}
  \xi(\br) = \exp[\frac{\ri}{2}\bm{\tau}\cdot\hat{\br}\,F(r)]\,.
\end{equation}
For the $\rho$ mesons, we use the Wu-Yang-'t~Hooft-Polyakov form as
\begin{equation}
  \label{eq:rho}
  \rho^{a, \mu=0}(\br) = 0\,,
  \qquad
  \rho^{a, \mu=i}(\br) = \varepsilon^{ika} \, \hat{r}^k \, \frac{G(r)}{gr}\,.
\end{equation}
For $\rho^{a,\mu=0}$, we sometimes write it as the three vector, $\brho^0$, with respect to the isospin index.
We note that $G(r)$ in the above form has nothing to do with the form factor, $\calG(t)$, in the previous section.  For the isosinglet vector meson, we take a purely temporal profile as
\begin{equation}
  \label{eq:omega}
  \omega^\mu(\br) = \omega(r) \, \delta^{\mu 0}\,,
\end{equation}
which is consistent with Eq.~\eqref{eq:LWZW} with $\omega_\mu$ couples to the baryon density, $B^0$, for a static soliton.

\begin{figure}
    \centering
    \includegraphics[width=0.6\linewidth]{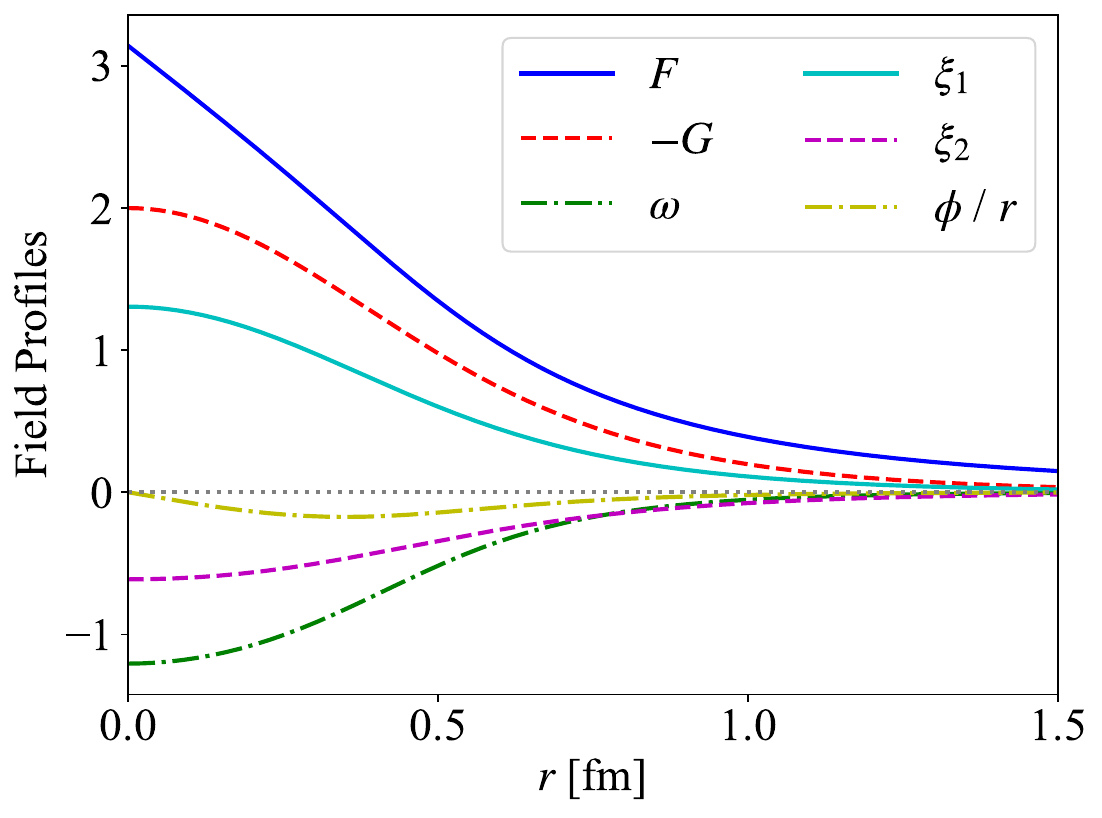}
    \caption{Numerical solutions of $F(r)$ (blue solid curve), $G(r)$ (red dashed curve), $\omega(r)$ (green dot-dashed curve), $\xi_1(r)$ (cyan solid line), $\xi_2(r)$ (magenta dotted curve), and $\phi(r)/r$ (yellow dot-dashed line).  Only $\omega(r)$ is rescaled with $f_\pi$ to be dimensionless.}
    \label{fig:field_profile}
\end{figure}

The equations of motion for $F(r)$, $G(r)$, and $\omega(r)$ are derived from the energy functional, and the resulting equations are
\begin{align}
  F^{\prime\prime} &= -\frac{2}{r} F^\prime - \frac{3g}{4\pi^2f_\pi^2r^2} \omega^\prime\sin^2F + \frac{1}{r^2} \left[4(G + 1)\sin F - \sin2F\right] + m_\pi^2 \sin F \,,\\
  G^{\prime\prime} &= \frac{1}{r^2} G(G + 1)(G + 2) + 2g^2 f_\pi^2\left(G + 1 - \cos F\right) \,,\\
  \omega^{\prime\prime} &= -\frac{2}{r}\omega^\prime + 2g^2f_\pi^2\omega
- \frac{3g}{4\pi^2r^2}F^\prime \sin^2 F\,.
\end{align}
The boundary conditions for a unit-winding soliton are $F(0)=\pi$ and $F(\infty)=0$ as usual.  Furthermore, to keep the energy finite,
\begin{align}
  G(0)=-2\,,\qquad G(\infty)=0\,,
  \qquad
  \omega^\prime(0)=0\,, \qquad \omega(\infty)=0\,,
\end{align}
should be imposed.  We numerically solve these equations to obtain the classical profile functions, $F(r)$, $G(r)$, and $\omega(r)$, as shown in Fig.~\ref{fig:field_profile}.

%--- figure ---%
%--- figure ---%

%%%%%
\subsection{Collective rotation with vector mesons}
\label{sec:quantization}

To describe nucleon states with definite spin, $J$, and isospin, $I$, we need to quantize the rotational zero modes of the $B=1$ soliton.
We employ the standard adiabatic $SU(2)$ rotation, i.e.,
\begin{equation}
  \label{eq:collective_rotation}
  U(\br,t) = A(t) \, U(\br) \, A^\dagger(t)\,,
\qquad
\bm\tau\cdot\bm\rho^{i}(\bm r,t)
= A(t)\bm\tau\cdot\bm\rho^{i}(\bm r)A^\dagger(t)\,,
\end{equation}
where $A(t)\in SU(2)$ is the collective coordinate.
The angular velocity vector, ${\bm K}$, in the body-fixed frame is defined by
\begin{equation}
  \label{eq:angvel_def}
  \btau\cdot\bm K
  = -\ri A^\dagger(t) \dot A(t)\,, 
\end{equation}
where $\dot{A}(t)=\mathrm{d}A(t)/\mathrm{d}t$.
%so that, in terms of the more common convention,
%$A^\dagger \dot A = \ri\bm\tau\cdot\bm\Omega/2$, one has $\bm\Omega=2\bm K$.
In the adiabatic expansion, we retain terms up to $\calO(K^2)$ in the collective Lagrangian
and solve the induced-field equations at $\calO(K)$.

The rotating soliton induces additional field components.  Not only $B^0$ but $B^i \propto (\bK \times \br)^i$ becomes nonzero, and then $\omega^i\neq 0$ is concluded from the WZW term in Eq.~\eqref{eq:LWZW}.  Therefore, we should parametrize $\omega^i$ as
\begin{align}
  \label{eq:omega_induced}
  \bm\omega(\br,t) = \frac{\phi(r)}{r} \bK \times \hat{\br}\,.
\end{align}
From the coupling in Eq.~\eqref{eq:L1}, the time derivative costs the kinetic energy, and the energy minimum is realized with $\brho^0 \neq 0$ where $\brho^0$ is the iso-vector with the Lorentz index $\mu=0$.  In other words, the time dependence of the collective rotation gives rise to isovector charge densities of $\calO(K)$, and the corresponding Gauss-law constraints require a nonvanishing $\brho^0 \sim \calO(K)$.  We thus parametrize $\brho^0$ as
\begin{align}
  \label{eq:rho0_induced}
  \btau\cdot\brho^{0}(\br,t)
  = \frac{2}{g}A(t)\, \btau\cdot\Bigl[ \bK\,\xi_1(r) + \hat{\br}(\bK\cdot\hat{\br})\,\xi_2(r) \Bigr]\, A^\dagger(t)\,.
\end{align}
Here, the radial profile functions, $\xi_1(r)$, $\xi_2(r)$, and $\phi(r)$, are determined by the equations of motion with the rotating background.

%%%%%
\subsection{Collective Lagrangian and induced meson profiles}
\label{sec:collective_lagrangian}

Substituting the parametrized fields in Sec.~\ref{sec:quantization} into the Lagrangian density
and expanding it in terms of $K$, the spatially integrated Lagrangian takes the rigid-rotor form as
\begin{equation}
    L = \int \dd[3]{x} \Lag_{\text{model}} = -E[F, G, \omega] + \Theta[F, G, \omega; \xi_1, \xi_2, \phi]\tr\bigl( \dot{A} \dot{A}^\dagger\bigr)\,,
\end{equation}
where $E$ is the static (non-rotating) soliton energy and $\Theta$ is the moment of inertia.
%The relation between $\tr(\dot A\dot A^\dagger)$ and the angular-momentum operator will be specified after quantization in Sec.~\ref{sec:quantized_rotor}.
The moment of inertia can be written as
\begin{equation}
    \Theta = 4\pi \int_0^\infty \dd{r} \Lambda(r)
\end{equation}
with the integrand given by
\begin{equation}
\begin{split}
    \Lambda(r) &= \frac{2}{3}f_\pi^2r^2\left[\sin^2F + 2(\xi_1 - 1 + \cos F)^2 + (\xi_1 + \xi_2)^2\right]  \\
    &\qquad + \frac{2}{3g^2}\left\{[G(1 - \xi_1) + \xi_2]^2 + [G(1 - \xi_1 - \xi_2) - \xi_2]^2\right\} + \frac{1}{3g^2}\left[2r^2{\xi_1^\prime}^2 +  r^2\left(\xi_1^\prime + \xi_2^\prime\right)^2\right] \\
    &\qquad - \frac{1}{3}f_\pi^2g^2\phi^2 - \frac{1}{6}{\phi^\prime}^2 - \frac{1}{3}\frac{\phi^2}{r^2} + \frac{1}{2\pi^2}g\phi F^\prime\sin^2F\,,
\end{split}
\end{equation}
where the prime denotes the radial derivative.
%$m\coloneqq m_\omega = m_\rho = \sqrt{2}f_\pi g$.

The equations of motion for $\xi_1(r)$, $\xi_2(r)$, and $\phi(r)$ are derived from the variational principle on the collective Lagrangian:
\begin{equation}
\begin{split}
    \xi_1^{\prime\prime} &= - 2\frac{\xi_1^\prime}{r} + 2f_\pi^2g^2(\xi_1 - 1 + \cos F) - \frac{G[G(1 - \xi_1) + \xi_2] + (G + 2)\xi_2}{r^2}\,,\\
    \xi_2^{\prime\prime} &= - 2\frac{\xi_2^\prime}{r} + 2f_\pi^2g^2(\xi_2 + 1 - \cos F) - \frac{G[G(1 - \xi_1 - \xi_2) - \xi_2] - (G + 2)^2\xi_2 - (G + 2)\xi_2}{r^2}\,,\\
    \phi^{\prime\prime} &= 2\frac{\phi}{r^2} + m^2\phi - \frac{3}{2\pi^2}gF^\prime\sin^2 F\,.
\end{split}
\end{equation}
Because these induced fields should be regular at the origin and vanish at spatial infinity, they should satisfy the following boundary conditions:
\begin{equation}
    \xi_i^\prime(0) = \xi_i(\infty) = 0\,,
    \qquad
    \phi(0) = \phi(\infty) = 0\,,
\end{equation}
where $i=1, 2$.  We numerically solve these equations for the fixed classical background profiles, $F(r)$, $G(r)$, and $\omega(r)$, and the determined profiles of $\xi_1(r)$, $\xi_2(r)$, and $\phi(r)$ are overlaid in Fig.~\ref{fig:field_profile}.

Note that we first minimize only $E[F, G, \omega]$ and subsequently add the rotational correction; the resulting field configuration is not an extremum of the full energy functional.  Equivalently, the Euler–Lagrange equation appropriate to the rotationally corrected functional is not satisfied.  Since the local conservation law, $\partial_\mu T^{\mu\nu}=0$, is an on-shell identity, this mismatch shows up as an apparent violation of EMT conservation in the static densities.  Concretely, the rotationally corrected pressure and shear-force densities need not obey the standard conservation constraints, such as the differential relation between $p(r)$ and $s(r)$ and the von~Laue condition.

%%%%%
\subsection{Quantization of the collective rotation}
\label{sec:quantized_rotor}

In the above description, the collective rotation is characterized by the angular velocity, $\bK$, which is a classical number.  The canonical quantization is a procedure to promote it to the operator acting on the nucleon state.  In practice, this amounts to interpreting $\bK$ as an operator via the relation to the rotational generator; that is, $\bK$ is replaced by the total spin operator, $\hat{\bm J}$, divided by the moment of inertia, i.e., $\bK \to \hat{\bm J}/(2\Theta)$.  Accordingly, the operator form of the rotational energy is
\begin{equation}
    \tr\bigl(\dot{A}\dot{A}^\dagger\bigr) = 2\bK^2 ~~\to~~ \frac{\hat{\bm{J}}^2}{2\Theta^2}\,.
\end{equation}
For the state with $J=1/2$, the eigenvalue of $\hat{\bm J}^2$ is $3/4$, and the rotational energy is
\begin{equation}
    \expval**{\frac{\hat{\bm{J}}^2}{2\Theta^2}}{J=1/2} = \frac{3}{8\Theta^2}\,.
\end{equation}
Note that, for the hedgehog soliton, the spatial and the isospin rotations are locked, and the collective quantization therefore generates only states with equal spin and isospin, $I=J$, e.g., the nucleon with $I=J=1/2$ and the $\Delta$ with $I=J=3/2$, etc.

Collective quantization is not merely a device to assign the quantum numbers to the baryon multiplets, but is essential for discussions on the EMT nucleon matrix elements.  At the classical level, the hedgehog form inevitably has vanishing momentum density, $T^{0i}=0$, so there is no way to access the local angular-momentum flow.  Time-dependent collective rotation introduces a momentum current, and only after projection onto a definite spin state such as $\ket{J=1/2}$, well-defined nucleon matrix elements, $\mel{p',s'}{T^{\mu\nu}}{p,s}$, are obtained.  Hence, the angular-momentum distribution and the associated EMT form factor, $J(t)$, arise after the quantization.

Moreover, in the present Skyrme model setup, the vector-meson sector carries an intrinsic spin density in a canonical decomposition.  Thus, contrary to intuition, the total angular momentum is not saturated by the orbital component associated with collective rotation.  It is therefore meaningful to track orbital and spin contributions from the mesonic degrees of freedom separately.  This is a crucial distinction from the pion-only Skyrme model without intrinsic spin brought in by the fields.

%%%%%%%%%%
\section{Results: comparison of the canonical and Belinfante EMTs}
\label{sec:emt_model}

We construct the EMT in our Skyrme-type model in two different schemes: the canonical EMT from the Noether formula and the Belinfante EMT that coincides with what is obtained by the metric variation.  In respective cases, we compute the EMT form factors from the model.

%%%%%
\subsection{Canonical EMT}
\label{sec:emt_canonical}

We start with the model Lagrangian density, $\Lag_{\text{model}}$, in Sec.~\ref{sec:model}.  Substituting this $\Lag_{\text{model}}$ into the canonical EMT in Eq.~\eqref{eq:Tcan_def}, we find that each Lagrangian part leads to
$T_{\text{can}}^{\mu\nu} = T_{\pi, \text{can}}^{\mu\nu} + T_{1, \text{can}}^{\mu\nu} + T_{V, \text{can}}^{\mu\nu} + T_{\text{WZW}, \text{can}}^{\mu\nu} + T_{m, \text{can}}^{\mu\nu}$,
where
\begin{align}
\label{eq:Tcan_model_detail}
    &T_{\pi, \text{can}}^{\mu\nu} = \frac{1}{4}f_\pi^2\tr(\partial^\mu U\partial^\nu U^\dagger + \partial^\nu U\partial^\mu U^\dagger) - g^{\mu\nu}\Lag_\pi\,, \\
    &T_{1, \text{can}}^{\mu\nu} = -f_\pi^2\tr(D^\mu\xi\cdot\xi^\dagger+D^\mu\xi^\dagger\cdot\xi)\left(\partial^\nu\xi\cdot\xi^\dagger+\partial^\nu\xi^\dagger\cdot\xi\right) - g^{\mu\nu}\Lag_1\,, \\
    &T_{V, \text{can}}^{\mu\nu} = -\frac{2}{g^2}\tr({F^\mu}_\lambda \partial^\nu V^\lambda) - g^{\mu\nu}\Lag_{V}\,, \\
    &T_{\text{WZW}, \text{can}}^{\mu\nu} = \frac{3g}{16\pi^2}\varepsilon^{\alpha\beta\gamma\mu}\omega_\alpha\tr(U^\dagger\partial_\beta UU^\dagger\partial_\gamma UU^\dagger\partial^\nu U) - g^{\mu\nu}\Lag_\text{WZW}\,, \\
    &T_{m, \text{can}}^{\mu\nu} = - g^{\mu\nu}\Lag_{m}\,.
\end{align}
In practice, we evaluate $T^{\mu\nu}_{\rm can}$ in the soliton form with vector mesons induced by classical rotation first, and then perform collective-coordinate quantization as described in Sec.~\ref{sec:quantization}.  The corresponding analytical expressions for $T_{\text{can}}^{\mu\nu}$ are summarized in Appendix~\ref{app:formula_Tcan}.

%%%%%
\subsection{Belinfante EMT}
\label{sec:emt_Belinfante}

To find the Belinfante EMT in our model, it would be easier to calculate the Hilbert EMT as a result of the metric variation, i.e.,
\begin{equation}
    T^{\mu\nu}_\text{Bel} = - 2\pdv{\Lag_\text{model}}{g_{\mu\nu}} - g^{\mu\nu}\Lag_\text{model}\,.
    \label{eq:THilbelt_model}
\end{equation}
Since $g_{\mu\nu}$ is symmetric with respect to $\mu$ and $\nu$, the Belinfante EMT,
$T_{\text{Bel}}^{\mu\nu} = T_{\pi, \text{Bel}}^{\mu\nu} + T_{1, \text{Bel}}^{\mu\nu} + T_{V, \text{Bel}}^{\mu\nu} + T_{\text{WZW}, \text{Bel}}^{\mu\nu} + T_{m, \text{Bel}}^{\mu\nu}$,
is also symmetric, where $T_{\pi, \text{Bel}}^{\mu\nu} = T_{\pi, \text{can}}^{\mu\nu}$, $T_{m, \text{Bel}}^{\mu\nu} = T_{m, \text{can}}^{\mu\nu}$, $T_{\text{WZW}, \text{Bel}}^{\mu\nu} = 0$ and
\begin{align}
\label{eq:Tbel_model_detail}
    T_{1, \text{Bel}}^{\mu\nu} &= -f_\pi^2 \tr[ (D^\mu\xi\cdot\xi^\dagger+D^\mu\xi^\dagger\cdot\xi) (D^\nu\xi\cdot\xi^\dagger+D^\nu\xi^\dagger\cdot\xi) ] - g^{\mu\nu} \Lag_1\,, \\
    T_{V, \text{Bel}}^{\mu\nu} &= -\frac{2}{g^2}\tr(F^\mu_{\;\;\lambda} F^{\nu\lambda}) - g^{\mu\nu}\Lag_{V}\,.
\end{align}
We can explicitly verify that this Belinfante EMT can be reproduced from the Belinfante symmetrization improvement from the canonical EMT, up to terms that vanish with the equations of motion.  It should be noted that the $T_{\text{WZW, Bel}}^{\mu\nu}=0$ because the Wess-Zumino-Witten term is topological and metric independent, so its variation with respect to $g_{\mu\nu}$ simply vanishes.  The corresponding analytical expressions for $T_{\text{Bel}}^{\mu\nu}$ are summarized in Appendix~\ref{app:formula_Tkin}.

% \subsection{Consistency checks}
% \label{sec:emt_checks}

% We perform several nontrivial checks on our implementation of both EMT constructions.

% \paragraph{(i) Conservation laws.}
% Using the classical equations of motion, we verify that $\partial_\mu T^{\mu\nu}=0$ holds for each construction
% (up to numerical accuracy), and that in the static limit the stress-tensor conservation reduces to the differential
% constraint
% \begin{equation}
% p'(r)+\frac{2}{3}\,s'(r)+\frac{2}{r}\,s(r)=0
% \label{eq:stress_conservation_check}
% \end{equation}
% for the pressure and shear distributions defined in Eq.~\eqref{eq:stress_decomp}.

% \paragraph{(ii) Charge normalization.}
% We confirm that the total energy reproduces the soliton mass and that the momentum form factor satisfies $A(0)=1$.
% In addition, for a spin-$1/2$ target, the total angular momentum is fixed by representation theory, implying
% \begin{equation}
% J(0)=\frac{1}{2},
% \label{eq:J0_half_check}
% \end{equation}
% which we use as a simple consistency check on the overall normalization of our matrix elements (independently of the
% pseudogauge choice).

% \paragraph{(iii) Mechanical stability.}
% We verify the von Laue condition,
% \begin{equation}
% \int_0^\infty dr\,r^2\,p(r)=0,
% \label{eq:vonLaue_check}
% \end{equation}
% which is required for a stable localized system, and we monitor its numerical accuracy for both EMT constructions.

\subsection{EMT form factors}
\label{sec:emt_extraction}

We extract the EMT form factors from off-forward nucleon matrix elements evaluated in the Breit frame ($\Delta^0=0$ and $t=-\bDelta^2$), using the conventions explained in Sec.~\ref{sec:general_framework}.  In the present work, we can only consider the total EMT, without quark and gluon decomposition, and use the minimal parametrization in terms of $A(t)$, $J(t)$, $D(t)$, and $\calG(t)$.  In the following, the subscripts ``can'' and ``Bel'' indicate that the form factors are extracted from the canonical EMT, $T^{\mu\nu}_\text{can}$, and from the Belinfante EMT, $T^{\mu\nu}_\text{Bel}$, respectively.

\begin{figure}
    \centering
    \includegraphics[width=0.6\textwidth]{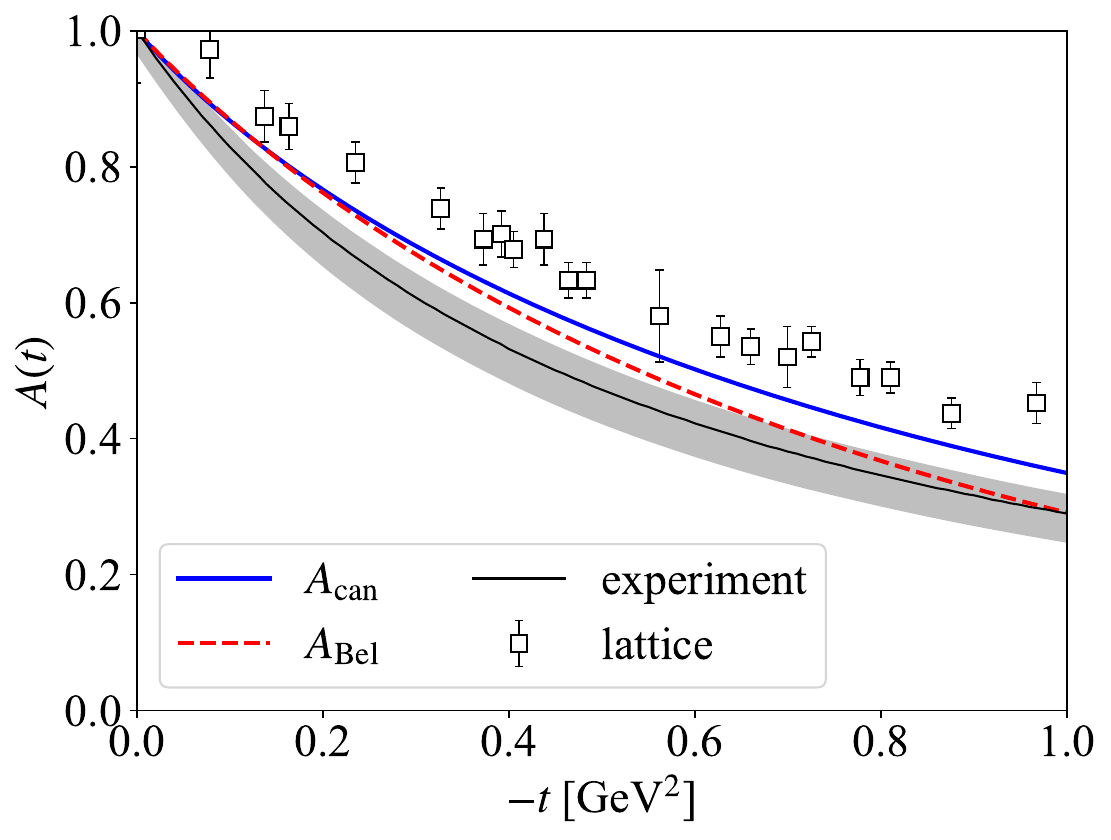}
    \caption{Momentum form factor, $A(t)$, extracted from $T^{\mu\nu}_\text{can}$ and $T^{\mu\nu}_\text{Bel}$.  In both cases, the normalization condition, $A(0)=1$, is satisfied within numerical accuracy.  For comparison, we also show the total lattice-QCD result of Ref.~\cite{Hackett:2023rif} and the phenomenological extraction based on near-threshold vector-meson photoproduction of Ref.~\cite{Wang:2023fmx}.}
    \label{fig:ff_A}
\end{figure}

\begin{figure}
    \centering
    \includegraphics[width=0.6\textwidth]{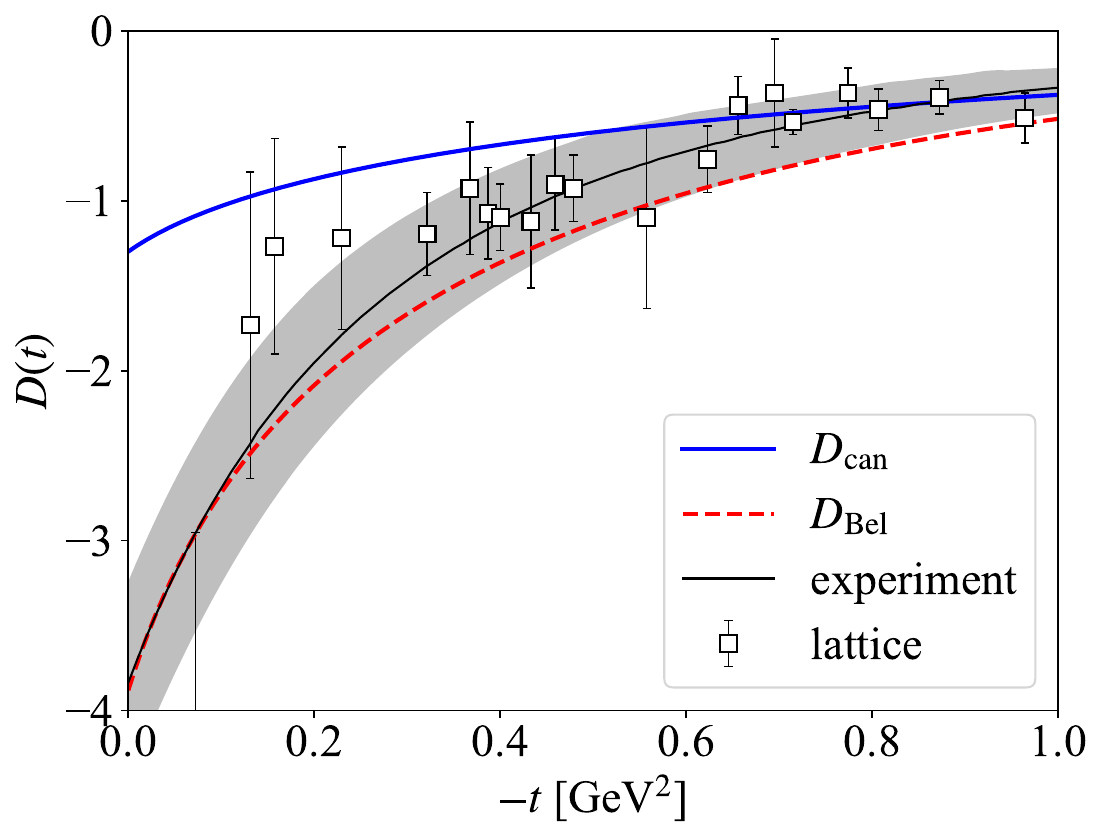}
    \caption{Mechanical form factor, $D(t)$, extracted from $T^{ij}_\text{can}$ and $T^{ij}_\text{Bel}$ using the static stress decomposition
    and Fourier-Bessel relations in Sec.~\ref{sec:inversion}.  For comparison, we also show the total lattice-QCD result of Ref.~\cite{Hackett:2023rif} and the phenomenological extraction based on near-threshold vector-meson photoproduction of Ref.~\cite{Wang:2023fmx}.}
    \label{fig:ff_D}
\end{figure}

\begin{figure}
    \centering
    \includegraphics[width=0.6\textwidth]{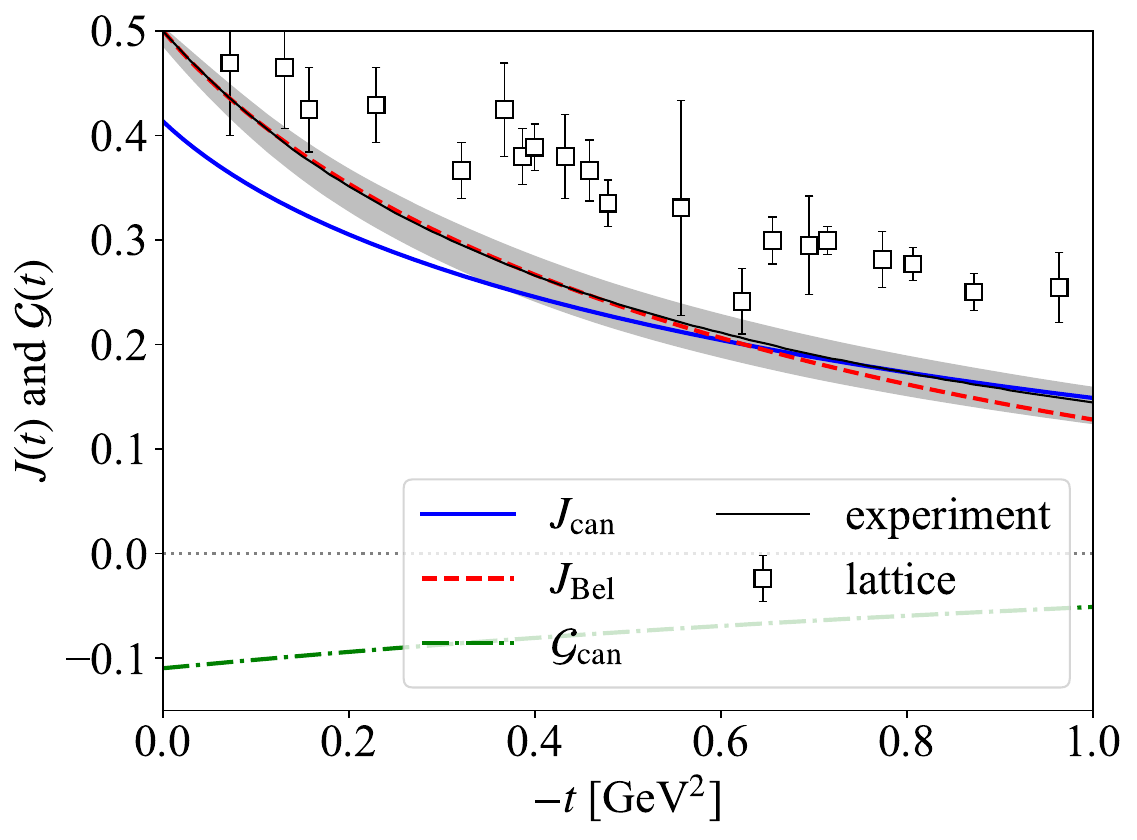}
    \caption{Angular-momentum form factor, $J(t)$, from the mixed components of the EMT\@.
    For the canonical EMT, we also quantify the antisymmetric contribution through an additional form factor, $\calG(t)$,
    which vanishes identically for the Belinfante EMT\@.  For comparison, we also show the total lattice-QCD result of Ref.~\cite{Hackett:2023rif} and the phenomenological extraction based on near-threshold vector-meson photoproduction of Ref.~\cite{Wang:2023fmx}.}
    \label{fig:ff_JG}
\end{figure}

We show the resulting form factors in Figs.~\ref{fig:ff_A}-\ref{fig:ff_JG}.  For comparison, we also display recent lattice-QCD results~\cite{Hackett:2023rif} and available phenomenological extractions from vector-meson photoproduction~\cite{Wang:2023fmx}.  Since these external results are usually formulated in terms of the Belinfante EMT, the direct comparison should be made with our Belinfante results.

Let us begin with the momentum form factor, $A(t)$, in Fig.~\ref{fig:ff_A}.  In the plot, $A_\text{can}(t)$ and $A_\text{Bel}(t)$ are obtained from the inversion formulas in Sec.~\ref{sec:inversion}.  As expected from translational invariance, both satisfy $A(0)=1$ within numerical accuracy.  We find that $A_\text{Bel}(t)$ shows a qualitatively similar decreasing behavior to the phenomenological extractions and lattice results.

%--- figure ---%
%--- figure ---%

Figure~\ref{fig:ff_D} shows the mechanical form factor, $D(t)$, obtained from the spatial stress tensor.  In our treatment, $D(t)$ based on the pressure profile in Eq.~\eqref{eq:D_from_p} does not exactly coincide with $D(t)$ based on the shear-force distribution in Eq.~\eqref{eq:D_from_s} due to the violation of the energy-momentum conservation caused by rotation.  In this work, $D(t)$ refers to that from the shear distribution, because it is numerically more stable.  In the forward limit, we find $D_\text{can}(0)=-1.30$ and $D_\text{Bel}(0)=-3.88$.  This comparison between $D_\text{can}(t)$ and $D_\text{Bel}(t)$ clearly illustrates how the pseudogauge choice could affect the local stress distribution and the extracted mechanical form factor.  In both cases, $D(0)$ is negative, as commonly discussed~\cite{Polyakov:2018zvc}.  We can see that $D_\text{Bel}(t)$ is in reasonably good agreement with the phenomenological extraction and lattice results.

%--- figure ---%
%--- figure ---%

The key message from Fig.~\ref{fig:ff_D} is that the local mechanical interpretation inferred from spatial stresses is not unique under pseudogauge transformations.  A prominent demonstration is that the forward limit value takes markedly different values, as mentioned above.
%$D_\text{can}(0)=-1.30$ and $D_\text{Bel}(0)=-3.88$.
As explained in Sec.~\ref{sec:partonic_relation}, the leading-twist DVCS constraints appear from the twist-2 GPD moments, and they only determine the symmetric leading-twist sector of the EMT, which does not resolve pseudogauge ambiguity in the spatial stress tensor.  Therefore, the DVCS data alone cannot fix which pseudogauge realization underlies a given three-dimensional pressure and shear-force reconstruction, and the mechanical interpretation in theory cannot avoid systematic uncertainty of the size indicated by $D_\text{can}(t)-D_\text{Bel}(t)$.

Finally, Fig.~\ref{fig:ff_JG} presents the angular-momentum form factor, $J(t)$, extracted from the mixed components.  For the Belinfante EMT, $T^{0i}_\text{Bel}=T^{i0}_\text{Bel}$ follows from the symmetric nature, and $\mathcal{G}(t)\equiv 0$ by construction.  For the canonical EMT, where $T^{0i}_\text{can}\neq T^{i0}_\text{can}$ in general, we extract $\calG(t)$ from the antisymmetric part as well as $J_\text{can}(t)$ from the symmetric combination, following the prescription in Sec.~\ref{sec:general_framework}.  At $t=0$, the normalization $J_\text{Bel}(0)=1/2$ reflects the spin-$1/2$ nature of the nucleon.  In close analogy with the mechanical form factor, the comparison of $J_\text{can}(t)$ and $J_\text{Bel}(t)$ provides an explicit estimate of the pseudogauge systematics in the off-forward angular-momentum sector.  The plot of $J_\text{Bel}(t)$ is in good agreement with the phenomenological extraction, while it is systematically smaller than the lattice results over the plotted range.  We make a remark here; when $J(t)$ is inferred phenomenologically through $J(t)=\frac{1}{2}\left[A(t)+B(t)\right]$ from twist-2 GPDs, the uncertainties in $A(t)$ and $B(t)$ would propagate directly into $J(t)$ at finite $t$.  In many phenomenological parameterizations, the exact forward-limit constraint for the total EMT, $A(0)=1$ and $B(0)=0$, is imposed, so that the experimental data primarily constrain the $t$-dependence rather than the overall normalization.

Unlike the Belinfante case, the canonical result finds $J_\text{can}(0)\neq 1/2$, not representing the total angular momentum.  It is worth emphasizing that the forward-limit normalization, $J_\text{Bel}(0)=1/2$, is enforced for the Belinfante EMT, for which the mixed components already encode the conserved total angular momentum.  In contrast, for the canonical EMT with $T^{0i}_\text{can}\neq T^{i0}_\text{can}$, $J_\text{can}(t)$ extracted from the symmetric combination need not, by itself, coincide with the total spin at $t=0$.  In our parametrization, the OAM charge in the canonical pseudogauge becomes $L_\text{can}=J_\text{can}(0)+\frac{1}{2}\mathcal{G}_\text{can}(0)=0.36$.  As shown in Sec.~\ref{sec:results_spin}, we can estimate the spin angular-momentum contribution in the same pseudogauge, and it turns out to be $S_\text{can}=0.14$.  In this way, we can confirm:
\begin{equation}
    L_\text{can} + S_\text{can} = \frac{1}{2}\,,
\end{equation}
within numerical accuracy.

We note that although $\calG(t)$ is sometimes interpreted as something related to the axial-vector form factor, $G_A(t)$, this identification applies to the kinetic Ji tensor only~\cite{Lorce:2015lna, Won:2025dgc}.  In our case, $\calG$ is defined from the antisymmetric part of the canonical EMT, and therefore, such a direct correspondence with the axial-vector form factor does not necessarily hold.

%--- figure ---%
%--- figure ---%

%%%%%%%%%%
\section{Results: Spin current tensor}\label{sec:results_spin}

This section presents the results of the spin current tensor in our model and the discussions of the local angular momentum decomposition.  Throughout this section, we consider polarized nucleon states with $s'=s$ and quantify the spin along the $z$ axis.

%%%%%
\subsection{Angular momentum current in the model}
\label{sec:spin_current_model}

For a general $\Lag_\text{model}$ depending on fields $\phi_a$ and their first derivatives, the canonical spin current obtained from the response to an infinitesimal Lorentz transformation can be written as
\begin{equation}
    S_\text{can}^{\lambda\mu\nu} = \sum_a \, \pdv{\Lag_\text{model}}{(\partial_\lambda \phi_a)}\,(\Sigma^{\mu\nu}\phi)_a
\label{eq:SpinCanDef_V}
\end{equation}
with $\Sigma^{\mu\nu}$ the Lorentz generator in the representation of $\phi_a$.
Since $\Sigma^{\mu\nu}$ is antisymmetric, we see $S_\text{can}^{\lambda\mu\nu}=-S_\text{can}^{\lambda\nu\mu}$.
In our model, $S_\text{can}^{\lambda\mu\nu}$ receives contributions only from vector-meson fields, $V^{\mu}$, i.e.,
\begin{equation}
    S_\text{can}^{\lambda\mu\nu} = - \frac{2}{g^2}\tr(F^{\lambda\mu}V^\nu - F^{\lambda\nu}V^\mu)\,.
\end{equation}
The full analytical expressions are found in App.~\ref{app:formula_Scan}.  The Belinfante-improved description corresponds to the pseudogauge choice in which the spin current is absorbed into the EMT, i.e., $S_\text{Bel}^{\lambda\mu\nu}=0$.

The formulas for the conserved angular momentum current, $J^{\lambda\mu\nu}$, and its orbital part, $L^{\lambda\mu\nu}$, are deduced from Eq.~\eqref{eq:LxT}, in which the OAM current is obtained from $T_\text{can}^{\mu\nu}$, and the similar construction for $T_\text{Bel}^{\mu\nu}$ leads to the total angular momentum current.  In what follows, we focus on the charge density of the angular momentum current tensor, i.e., $J^{0ij}$ and $J^{00i}$, which correspond to the angular momentum density and the boost density, respectively, in the chosen pseudogauge.

It is convenient to introduce the angular momentum three-vectors as
\begin{equation}
    L^m=\frac{1}{2}\epsilon^{mij}L^{0ij}\,,
    \qquad
    S^m=\frac{1}{2}\epsilon^{mij}S^{0ij}\,,
    \qquad
    J^m=L^m+S^m\,,
\label{eq:VectorAM_V}
\end{equation}
and to study their radial profiles for the polarized soliton.

%%%%%
\subsection{Polarized distributions}
\label{sec:spin_results}

We now present the spatial distributions of the orbital and spin contributions for a polarized nucleon.  All densities shown below are obtained in the Breit frame.

\begin{figure}
    \centering
    \includegraphics[width=0.49\textwidth]{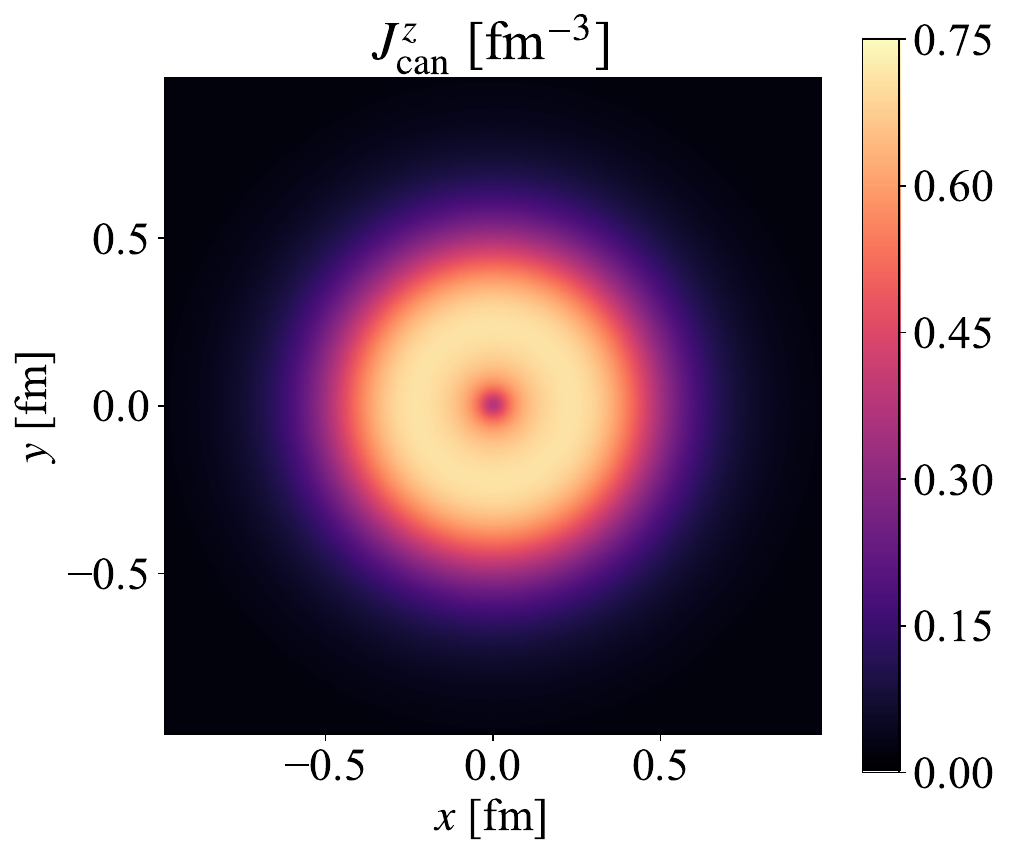}
    \includegraphics[width=0.49\textwidth]{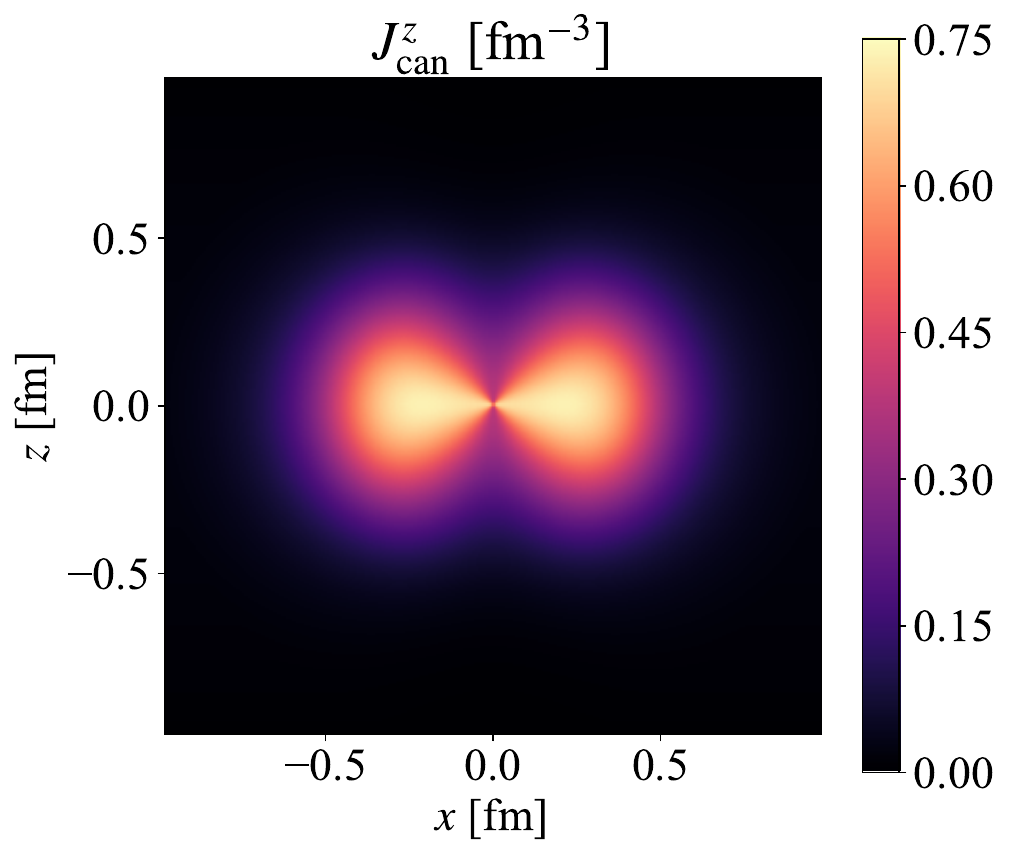}
    \caption{Canonical total angular-momentum density, $J^z_\text{can}$, for a nucleon polarized along $\hat{\bm z}$.
    Left: transverse plane ($z=\SI{0.05}{fm}$). Right: longitudinal plane ($y=0$). These two slices visualize the characteristic geometry implied by rotational symmetry along the polarization axis.}
    \label{fig:Jz_can_xy}
\end{figure}

\begin{figure}
    \centering
    \includegraphics[width=0.49\textwidth]{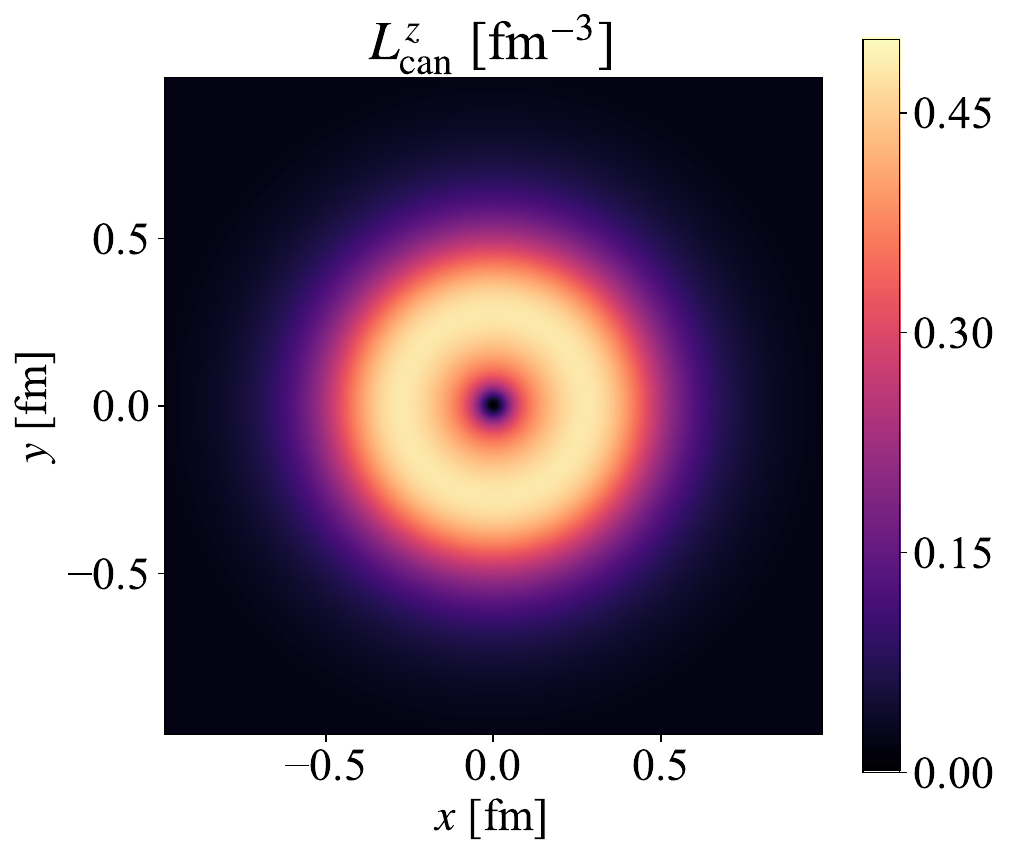}
    \includegraphics[width=0.49\textwidth]{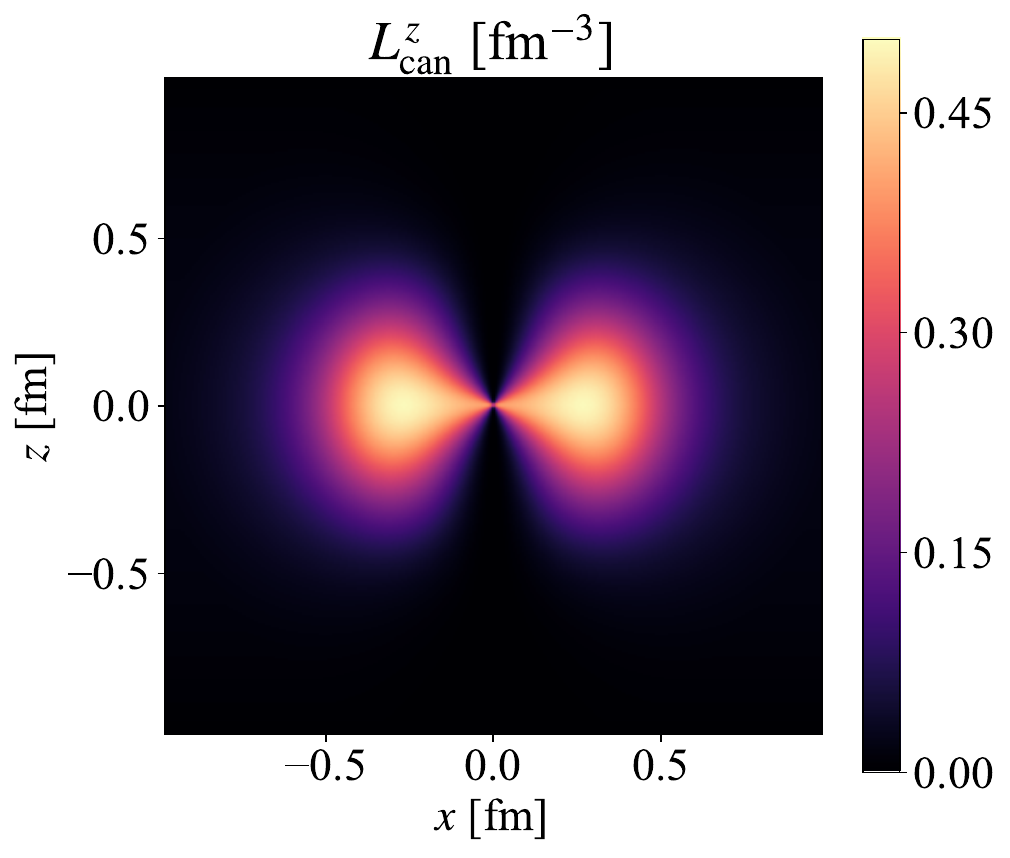}
    \caption{Canonical OAM density, $L^z_\text{can}$, for a nucleon polarized along $\hat{\bm z}$.
    Left: transverse plane ($z=\SI{0.05}{fm}$). Right: longitudinal plane ($y=0$).}
    \label{fig:Lz_can_xy}
\end{figure}

\begin{figure}
    \centering
    \includegraphics[width=0.49\textwidth]{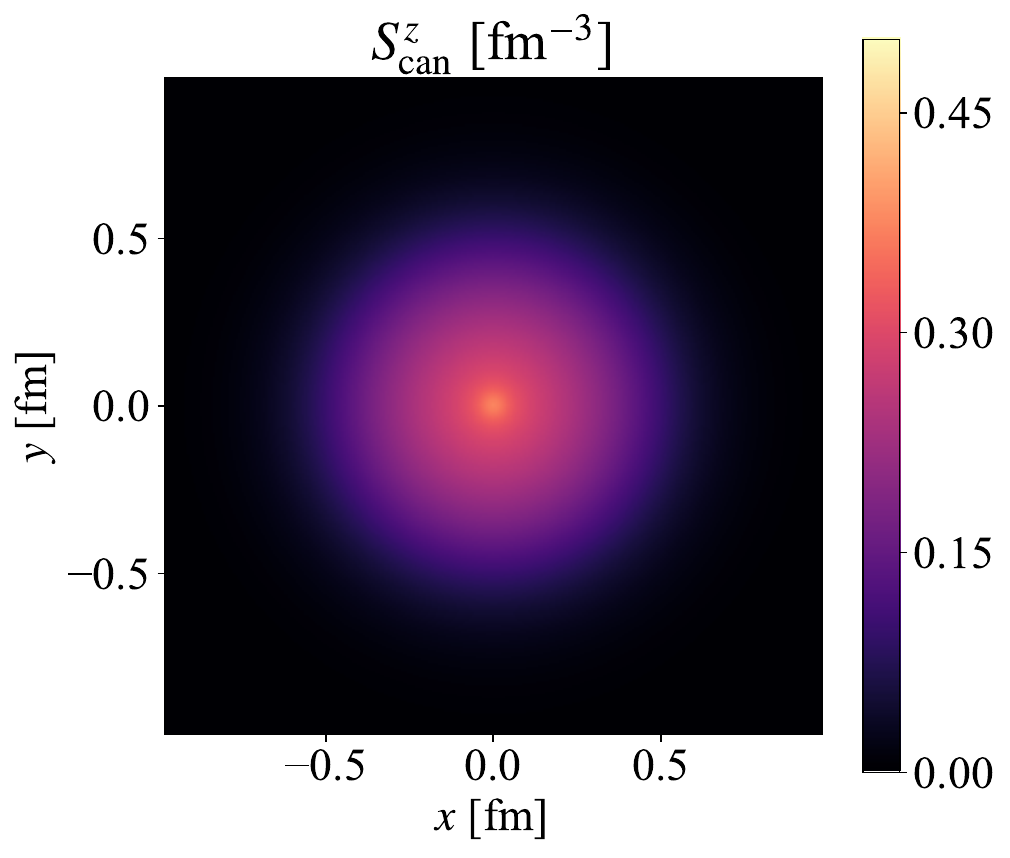}
    \includegraphics[width=0.49\textwidth]{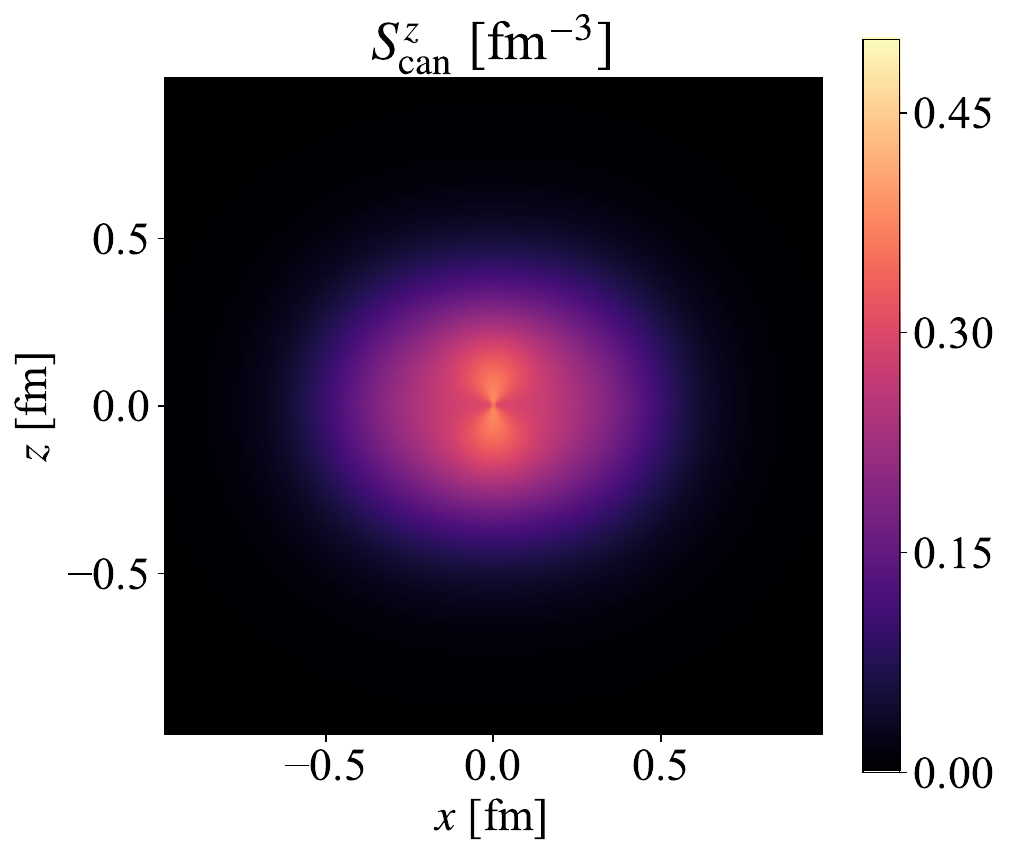}
    \caption{Canonical spin density, $S^z_\text{can}$, for a nucleon polarized along $\hat{\bm z}$.
    Left: transverse plane ($z=\SI{0.05}{fm}$). Right: longitudinal plane ($y=0$).}
    \label{fig:Sz_can_xy}
\end{figure}

Figure~\ref{fig:Jz_can_xy} shows the $z$ component of the total angular-momentum density in the canonical pseudogauge, $J^z_\text{can}$, plotted in the transverse plane at $z=\SI{0.05}{fm}$ (left) and the longitudinal plane at $y=0$ (right).  For comparison, Figs.~\ref{fig:Lz_can_xy} and \ref{fig:Sz_can_xy} display the corresponding canonical orbital and spin densities, $L^z_\text{can}$ and $S^z_\text{can}$, respectively, in the same planes.  These distributions clearly visualize how the conserved total angular momentum is locally distributed into the orbital and spin contributions in the canonical formulation.

%--- figure ---%

%--- figure ---%

\begin{figure}
    \centering
    \includegraphics[width=0.49\textwidth]{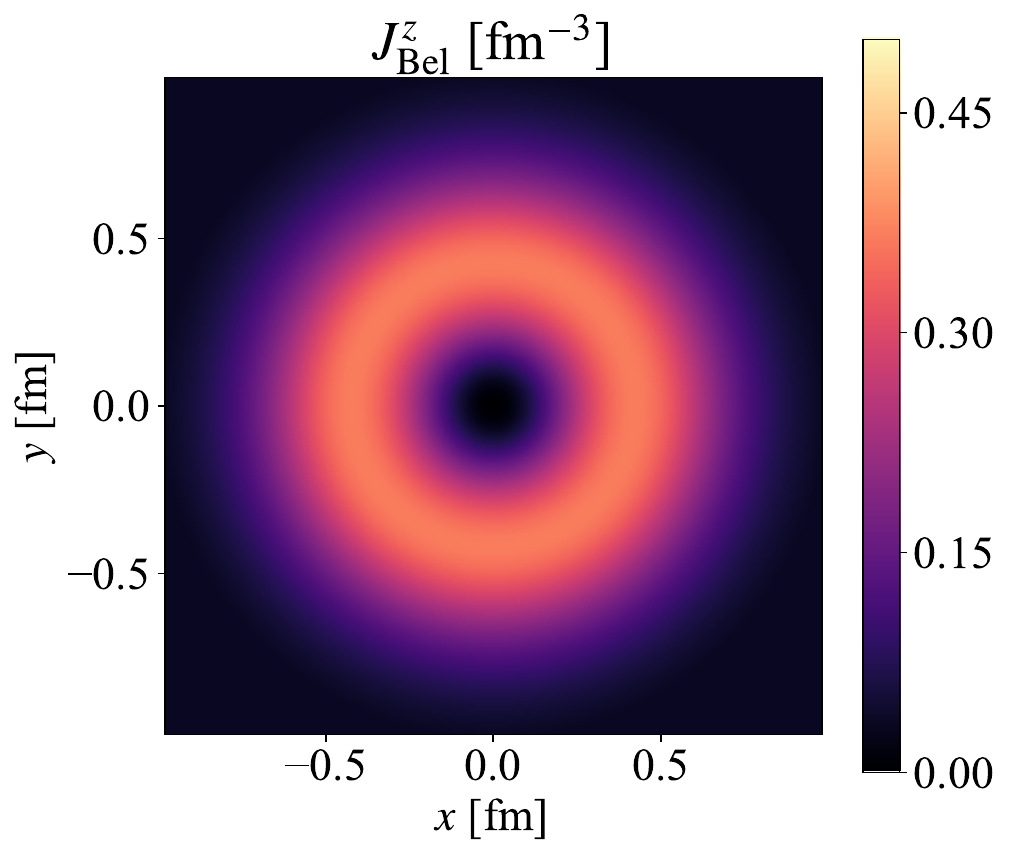}
    \includegraphics[width=0.49\textwidth]{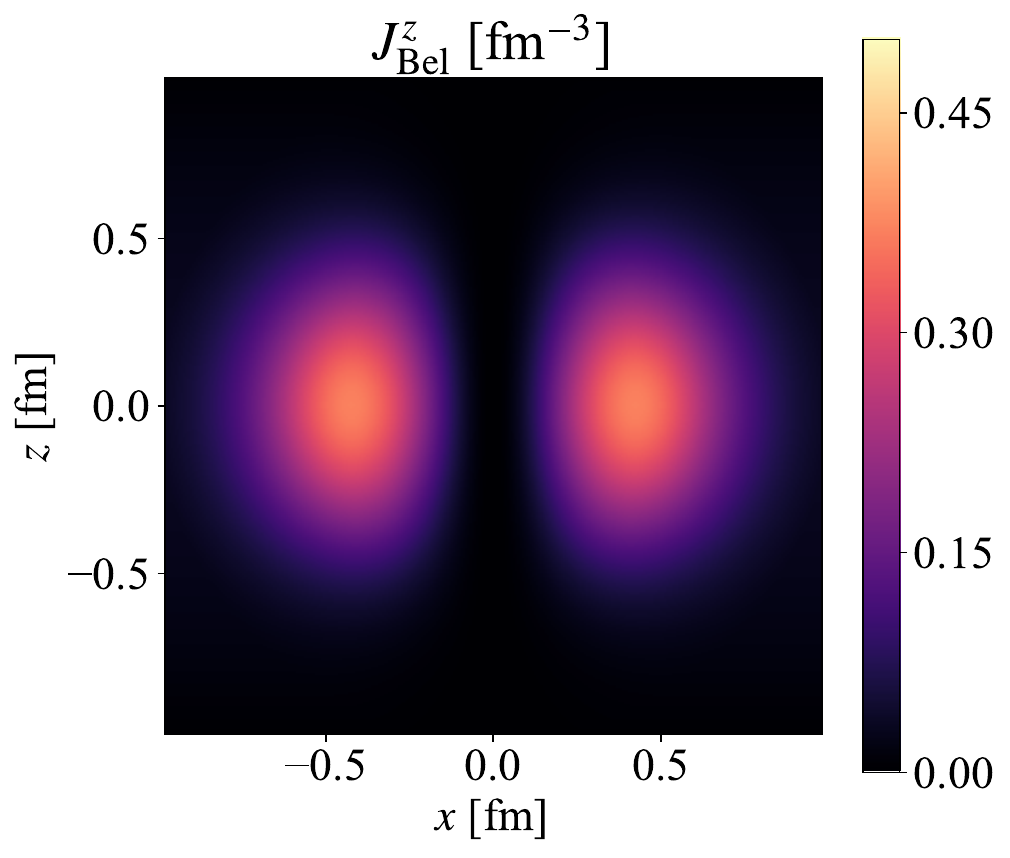}
    \caption{Belinfante angular-momentum density, $J^z_\text{Bel}$, for a nucleon polarized along $\hat{\bm z}$.
    Left: transverse plane ($z=\SI{0.05}{fm}$). Right: longitudinal plane ($y=0$).}
    \label{fig:Lz_bel_planes}
\end{figure}

In the Belinfante pseudogauge, the spin current is absorbed into the EMT, so that the total angular-momentum density is encoded entirely in the orbital-like form, $x^iT^{0j}-x^jT^{0i}$, built from the Belinfante EMT\@.  Figure~\ref{fig:Lz_bel_planes} shows the corresponding density $J^z_\text{Bel}$ in two representative slices on the $xy$ plane at $z=\SI{0.05}{fm}$ and the $xz$ plane at $y=0$.  The transverse slice manifests the azimuthally symmetric profile, while the $xz$ slice highlights the expected suppression on the polarization axis dictated by the minimal tensor structure for a polarized, spherically symmetric configuration.  Since the Belinfante total angular momentum is written in the orbital-like form, $x^iT^{0j}-x^jT^{0i}$, its density vanishes at the center of the nucleon.  This behavior makes a sharp contrast with the canonical results in Fig.~\ref{fig:Jz_can_xy} in which the spin contribution clearly remains finite even at the center.

%--- figure ---%

\begin{figure}
    \centering
    \includegraphics[width=0.47\textwidth]{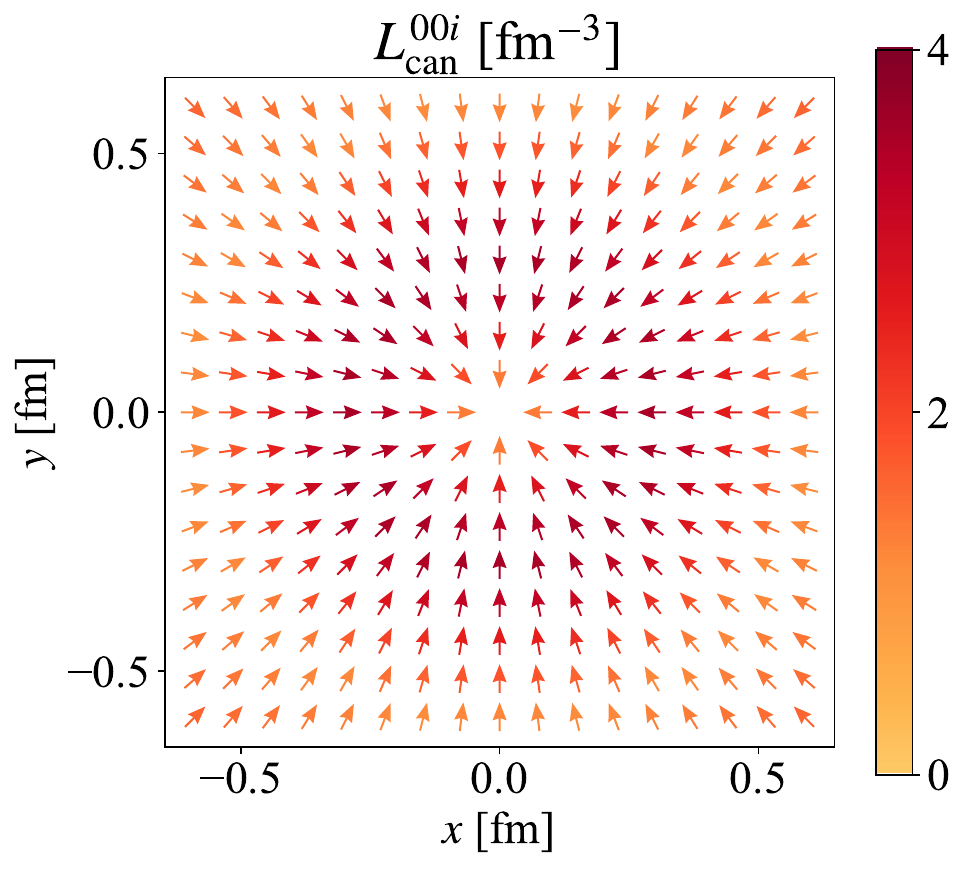}
    \includegraphics[width=0.49\textwidth]{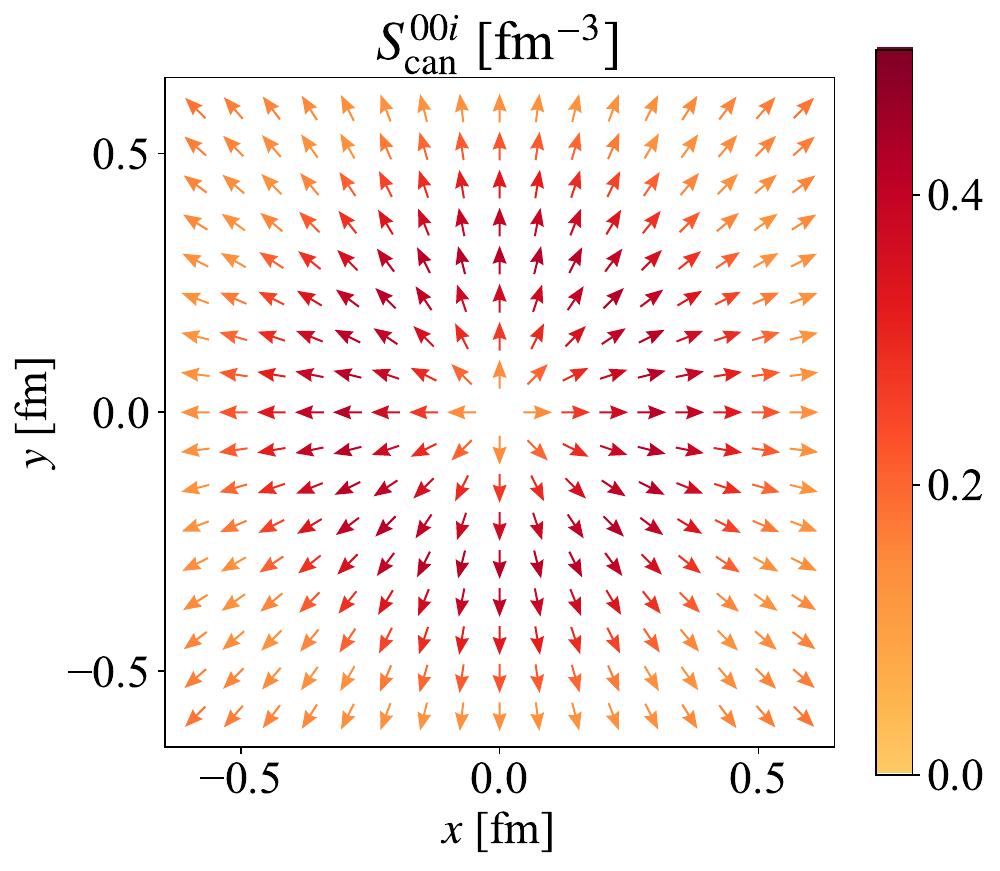}
    \caption{Canonical decomposition of the boost charge density into the orbital part $L^{00i}_\text{can}$ (left) and the spin part $S^{00i}_\text{can}$ (right).  The boost charge density is spherically symmetric, and only the distribution on the $z=0$ plane is shown.}
    \label{fig:boost_can_planes}
\end{figure}

Finally, Fig.~\ref{fig:boost_can_planes} presents the boost-related distributions in the transverse plane, comparing the orbital boost density $L^{00i}$ and the spin boost density $S^{00i}$ in the canonical one.

\begin{figure}
    \centering
    \includegraphics[width=0.47\textwidth]{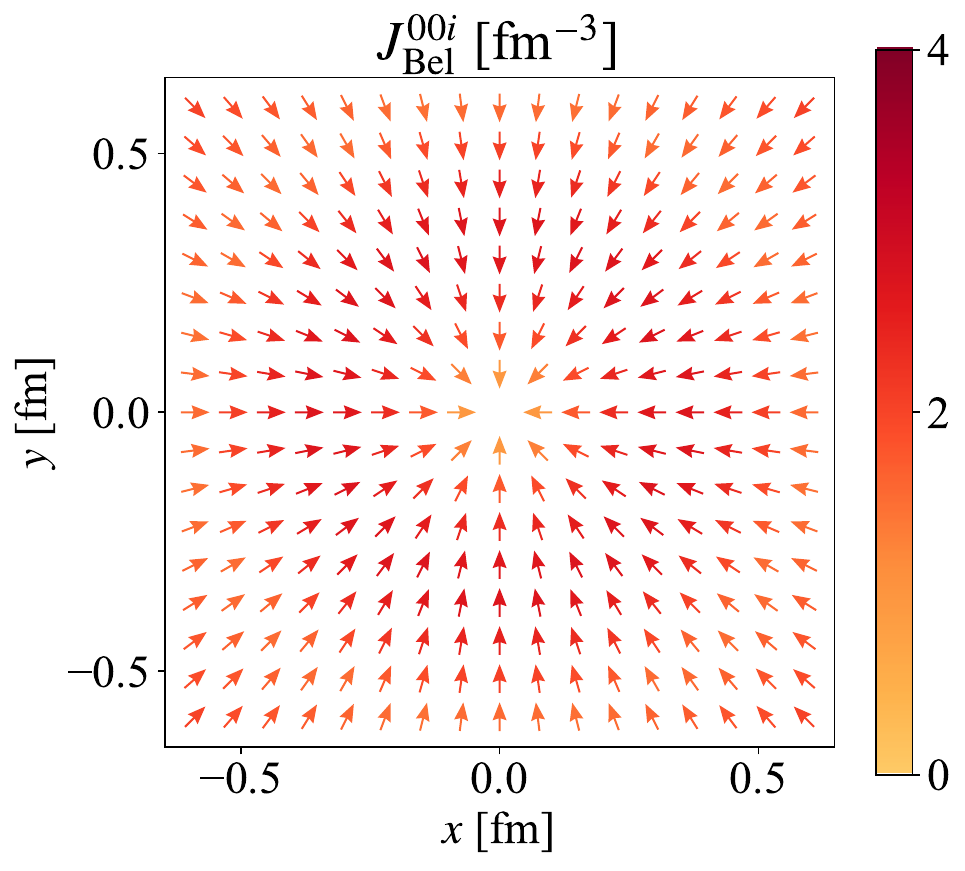}
    \caption{Belinfante boost charge density, $J^{00i}_\text{Bel}$.}
    \label{fig:boost_Bel_planes}
\end{figure}
%--- figure ---%

The orbital part of the boost densities in Figs.~\ref{fig:boost_can_planes} and \ref{fig:boost_Bel_planes} plots $L^{00i}=x^{0}T^{0i}-x^{i}T^{00}$, which reduces to $L^{00i}=-x^{i}T^{00}$ in the $x^0\to 0$ limit.  Hence, $L^{00i}$ represents the moment of the energy density distribution.  The spatial integral of $L^{00i}$ gives the charge; $L^{0i} = \int\dd[3]{x}L^{00i}= - \int \dd[3]{x}x^{i}T^{00}$, that is nothing but the center of energy.  In our setup, the nucleon is at rest, and we choose the origin at its center, so the total charge vanishes identically, $L^{0i}=0$.  This is also consistent with our numerical profiles in Figs.~\ref{fig:boost_can_planes} and \ref{fig:boost_Bel_planes}. The vector field exhibits a radial inward pattern, so that contributions from opposite directions cancel upon integration.

Although our numerical results are obtained for a static nucleon in its rest frame, so that there is no net momentum and no transverse polarization needs to be invoked, it is worth noting that the corresponding boost charge and the associated transverse center-of-energy shift play a central role in the impact-parameter picture~\cite{Burkardt:2002hr, Burkardt:2005hp} and in the long-standing discussions of angular-momentum decompositions of transversely polarized nucleons; see, e.g., Refs.~\cite{Leader:2011cr, Ji:2012sj, Ji:2012vj, Leader:2012ar, Harindranath:2012wn, Ji:2013tva, Harindranath:2013goa, Ji:2020hii, Guo:2021aik}.

As seen in Fig.~\ref{fig:boost_can_planes} qualitatively, the canonical spin contribution $S^{00i}_\text{can}$ also integrates to zero in our static setup, i.e.,
\begin{equation}
    \int \dd[3]{x}S^{00i}_\text{can}=0\,.
\end{equation}
It is useful to note that, in the gauge-invariant light-front constructions of the spin current, the would-be ``spin-boost'' component $S^{++i}$ becomes essentially trivial at the operator level.  Indeed, the quark-spin part is proportional to the axial structure $S^{\lambda\mu\nu}_{q}=\frac{1}{2}\epsilon^{\lambda\mu\nu\sigma}\bar\psi\gamma_\sigma\gamma_5\psi$, so $S^{++i}_{q}=0$ identically.  For gluons, one has $S^{\lambda\mu\nu}_{g}=F^{\lambda\nu}A^{\mu}_\text{phys}-F^{\lambda\mu}A^{\nu}_\text{phys}$, hence $S^{++i}_{g}=-F^{+i}A^{+}_\text{phys}$, which vanishes in the light-cone gauge $A^+_\text{phys}=0$~\cite{Hatta:2011zs}.  Therefore, $S^{++i}$ can be set to zero at the operator level, in contrast to the helicity-relevant component, $S^{+12}$, which is tied to $\Delta\Sigma$ and $\Delta G$ in the light-front partonic picture~\cite{Hatta:2011zs}.

% From the viewpoint of Poincar\'e symmetry, the clean intrinsic labels of a relativistic one-particle state are provided by the momentum $P^\mu$ together with the Pauli-Lubanski vector,
% \begin{equation}
%     W^\mu = \frac{1}{2}\epsilon^{\mu\nu\rho\sigma} P_\nu J_{\rho\sigma}\,.
% \end{equation}
% A particularly convenient spin quantum number is the helicity, i.e., the projection of spin along the direction of motion.  In contrast, separating this intrinsic spin information into a spin charge and a boost charge is not straightforward: the Lorentz generators $J^{\mu\nu}$ themselves are frame-dependent, and their decomposition into rotations and boosts depends on the choice of time slice and reference point.  In particular, the boost charge $J^{0i}$ can always be made to vanish by an appropriate choice of frame.
% For a localized system one may write
% \begin{equation}
% J^{0i}=\int d^{3}x\,(x^{0}T^{0i}-x^{i}T^{00})
% = x^{0}P^{i}-\int d^{3}x\,x^{i}T^{00}\,,
% \end{equation}
% so in the rest frame $P^{i}=0$ and with the origin chosen at the (energy) centroid at $x^{0}=0$ one has $J^{0i}=0$.
% Local densities such as $L^{00i}(\bm x)$ or $S^{00i}(\bm x)$ still exist on a fixed time slice,
% but precisely because $J^{0i}$ is not an invariant intrinsic label, these local ``boost densities'' do not
% admit a direct, frame-independent observational interpretation in the same way as helicity (or the total spin)
% does.

%%%%%%%%%%
\section{Conclusions}
\label{sec:conclusions}

We studied polarized EMT form factors and spatial spin densities in the nucleon in a field-theoretical soliton model, focusing on how pseudogauge freedom affects local interpretations.  Using a vector-meson extended Skyrme-type model and collective-coordinate quantization, we constructed both the canonical EMT, $T^{\mu\nu}_\text{can}$, and the Belinfante EMT, $T^{\mu\nu}_\text{Bel}$, within the same dynamical framework.  This setup is essential for addressing polarized structures: at the classical hedgehog level, the momentum density vanishes, so nontrivial mixed components and hence angular-momentum flow arise only after introducing time-dependentt collective rotation and projecting onto definite-spin nucleon states.

From off-forward nucleon matrix elements, we extracted a minimal set of form factors, $A(t)$, $J(t)$, and $D(t)$ for both pseudogauges, and, for the non-symmetric canonical tensor, the additional antisymmetric form factor $\mathcal{G}(t)$.  Translational invariance is respected numerically through $A(0)=1$ in both constructions.  In contrast, the mechanical sector exhibits a marked pseudogauge dependence: in our rotational scheme, the $D$-term determined from the shear-force distribution yields
$D_\text{can}(0)=-1.30$ and $D_\text{Bel}(0)=-3.88$.
This explicit difference demonstrates that mechanical interpretations based on spatial stress distributions are not unique under pseudogauge transformations.  Since leading-twist GPD constraints probe only the symmetric, light-front projected sector, they cannot by themselves resolve which pseudogauge underlies a three-dimensional reconstruction of pressure and shear forces.  Our comparison, therefore, provides a concrete model estimate of the size of the associated systematic ambiguity in the $t$-dependence and normalization of mechanical observables.

We also analyzed the angular-momentum sector at both the form-factor and density levels.  For the Belinfante tensor, the normalization $J_\text{Bel}(0)=1/2$ follows directly from the symmetry of the EMT for a spin-$1/2$ target.  For the canonical tensor, where $T^{0i}_\text{can}\neq T^{i0}_\text{can}$, the quantity $J_\text{can}(t)$ extracted from the symmetric combination does not need to satisfy $J_\text{can}(0)=1/2$ by itself; the antisymmetric contribution encoded in $\mathcal{G}(t)$ and the explicit spin-current contribution are required for a complete accounting.  In the canonical pseudogauge, we obtained an orbital contribution $L_\text{can}=J_\text{can}(0)+\frac{1}{2}\mathcal{G}_\text{can}(0)=0.36$, and a nonvanishing intrinsic-spin contribution $S_\text{can}=0.14$ originating from the vector-meson sector, confirming $L_\text{can}+S_\text{can}=1/2$ within numerical accuracy.  We further visualized the corresponding three-dimensional densities for a longitudinally polarized nucleon, highlighting the characteristic geometries implied by rotational symmetry: in the canonical construction, the conserved total angular momentum is locally deposited to the orbital and spin densities, whereas in the Belinfante construction it is entirely encoded in the orbital form.  Finally, we discussed boost-related densities and verified that the corresponding charges vanish in the nucleon rest frame.

The present framework is intended as a controlled laboratory to quantify and visualize pseudogauge effects in polarized EMT observables, rather than as a precision phenomenology tool.  Known limitations include the reduced axial coupling typical of this model and small violations of exact EMT conservation inherent in the semiclassical rotational treatment.  An important direction for future work is to refine the dynamics and to explore how pseudogauge-dependent information could be accessed through observables beyond leading twist (e.g., higher-twist GPD structures or transverse-momentum dependent/GTMD correlators), thereby strengthening the connection between EMT tomography and operator definitions of spin and mechanical structure.  It is also an intriguing problem to utilize the present model approach to study the force densities in Eq.~\eqref{eq:force_density}~\cite{Ji-private}, which can be pseudogauge insensitive and unambiguous observables, to disentangle genuinely physical features from prescription-dependent characterization of the EMT local properties.

%%%%%%%%%%
\begin{acknowledgments}
The authors thank
Daisuke~Fujii,
Xiangdong~Ji,
Mamiya~Kawaguchi,
Mitsuru~Tanaka,
Dong-Lin~Wang
for useful discussions and comments.
The authors thank the Yukawa Institute for Theoretical Physics (YITP) at Kyoto University. Discussions during the YITP school YITP-W-25-16 on ``YITP International School on EIC Physics'' were useful to complete this work.
This work was partially supported by Japan Society for the Promotion of Science
(JSPS) KAKENHI Grant No.\ 
22H05118 (K.F.) and
FoPM WINGS Program at The University of Tokyo (T.U.).
\end{acknowledgments}

%%%%%%%%%%
\appendix

%%%%%%%%%%
\section{Analytical expressions of the EMT in the model}

Explicit analytical expressions for the EMT in the vector-meson extended Skyrme-type model are listed here.  All formulas are evaluated for the rotating hedgehog Ansatz and expressed in terms of the static radial profiles $F(r)$, $G(r)$, and $\omega(r)$, together with the induced radial profiles $\xi_1(r)$, $\xi_2(r)$, and $\phi(r)$ generated by collective rotation.

%%%%%
\subsection{Canonical EMT}\label{app:formula_Tcan}

The canonical EMT, $T^{\mu\nu}_\text{can}$, is obtained by the Noether procedure applied to spacetime translations.  In general, it is not symmetric, and one therefore distinguishes $T^{0i}_\text{can}$ and
$T^{i0}_\text{can}$.  Below we list the explicit expressions for the energy density, $T^{00}_\text{can}$, the spatial stress tensor, $T^{ij}_\text{can}$, the energy flow $T^{0i}_\text{can}$, and the momentum density, $T^{i0}_\text{can}$:

\begin{equation}
\begin{split}
    T_{\text{can}}^{00} &= \frac{1}{2}f_\pi^2\left({F^\prime}^2 + 2\frac{\sin^2 F}{r^2}\right) + 2f_\pi^2\frac{(G + 1 - \cos F)^2}{r^2} - f_\pi^2 g^2\omega^2 + \frac{{G^\prime}^2}{g^2r^2} + \frac{G^2(G + 2)^2}{2g^2r^4} \\
    &\qquad - \frac{1}{2}{\omega^\prime}^2 + \frac{3g}{4\pi^2r^2}\omega F^\prime\sin^2 F + f_\pi^2m_\pi^2(1 - \cos F) \\
    &\qquad + \frac{1}{2\Theta^2}\Biggl\{f_\pi^2\sin^2F - 2f_\pi^2(\xi_1 - 1 + \cos F)(\xi_1 + 1 - \cos F) + \frac{1}{2}f_\pi^2g^2\frac{\phi^2}{r^2} \\
    &\qquad + \frac{[G(1 + \xi_1) - \xi_2][G(1 - \xi_1) + \xi_2]}{g^2r^2} - \frac{1}{g^2}{\xi_1^\prime}^2 + \frac{{\phi^\prime}^2}{4r^2}\Biggr\}\left[\bm{J}^2 - (\bm{J}\cdot\hat{\bm{r}})^2\right] \\
    &\qquad + \frac{1}{2\Theta^2}\Biggl\{-2f_\pi^2(\xi_1 + \xi_2)^2 + \frac{2[G(1 + \xi_1 + \xi_2) + \xi_2][G(1 - \xi_1 - \xi_2) - \xi_2]}{g^2r^2} \\
    &\qquad - \frac{1}{g^2}(\xi_1^\prime + \xi_2^\prime)^2 + \frac{\phi^2}{r^4}\Biggr\}\left(\bm{J}\cdot\hat{\bm{r}}\right)^2\,,
\end{split}
\end{equation}

\begin{equation}
\begin{split}
    T_{\text{can}}^{ij} &= - \frac{1}{2}f_\pi^2{F^\prime}^2\delta^{ij} + f_\pi^2\left({F^\prime}^2 - \frac{\sin^2 F}{r^2}\right)\hat{r}^i\hat{r}^j - 2f_\pi^2\frac{G(G + 1 - \cos F)}{r^2}\delta^{ij} \\
    &\qquad - 2f_\pi^2\frac{(1 - \cos F)(G + 1 - \cos F)}{r^2}\hat{r}^i\hat{r}^j + f_\pi^2 g^2\omega^2\delta^{ij} + \frac{1}{2}{\omega^\prime}^2(\delta_{ij} - 2\hat{r}_i\hat{r}_j) \\
    &\qquad - \frac{G^3(G + 2)}{2g^2r^4}\delta^{ij} - \frac{G^2(G + 2)}{g^2r^4}\hat{r}^i\hat{r}^j - \frac{{G^\prime}^2}{g^2r^2}(\delta^{ij} - 2\hat{r}^i\hat{r}^j) + \frac{GG^\prime}{g^2r^3}(\delta^{ij} - 3\hat{r}^i\hat{r}^j) \\
    &\qquad - f_\pi^2m_\pi^2(1 - \cos F)\delta^{ij} \\
    &\qquad + \frac{1}{2\Theta^2}\Biggl\{f_\pi^2\sin^2F\delta^{ij} + 2f_\pi^2(\xi_1 - 1 + \cos F)^2\delta^{ij} - \frac{1}{2}f_\pi^2g^2\frac{\phi^2}{r^2}\delta^{ij} \\
    &\qquad + \frac{\left[G(1 - \xi_1) + \xi_2\right]^2}{g^2r^2}\delta^{ij} + \frac{1}{g^2}{\xi_1^\prime}^2(\delta^{ij} - 2\hat{r}^i\hat{r}^j) - \frac{\phi^\prime}{4r^2}\left(\phi^\prime - \frac{2\phi}{r}\right)(\delta^{ij} - 2\hat{r}^i\hat{r}^j) \\
    &\qquad + \frac{3g}{4\pi^2r^2}\phi F^\prime\sin^2 F(\delta^{ij} - \hat{r}^i\hat{r}^j)\Biggr\}\left[\bm{J}^2 - (\bm{J}\cdot\hat{\bm{r}})^2\right] \\
    &\qquad + \frac{1}{2\Theta^2}\Biggl\{2f_\pi^2(\xi_1 + \xi_2)^2\delta^{ij} + \frac{2G(1 - \xi_1 - \xi_2)\left[G(1 - \xi_1 - \xi_2) - \xi_2\right]}{g^2r^2}\delta^{ij} \\
    &\qquad + \frac{1}{g^2}(\xi_1^\prime + \xi_2^\prime)^2(\delta^{ij} - 2\hat{r}^i\hat{r}^j) - \frac{2G(1 - \xi_1 - \xi_2)\xi_2}{g^2r^2}\hat{r}^i\hat{r}^j - \frac{2G(1 - \xi_1)\xi_2}{g^2r^2}\hat{r}^i\hat{r}^j \\
    &\qquad - \frac{2[G(1 - \xi_1) + \xi_2]\xi_2}{g^2r^2}(\delta^{ij} - 2\hat{r}^i\hat{r}^j) + \frac{\phi}{2r^3}\left(\phi^\prime - \frac{2\phi}{r}\right)\hat{r}^i\hat{r}^j - \frac{\phi\phi^\prime}{2r^3}\delta^{ij} \\
    &\qquad - \frac{3g}{4\pi^2r^2}\phi F^\prime\sin^2 F(\delta^{ij} - \hat{r}^i\hat{r}^j)\Biggr\}\left(\bm{J}\cdot\hat{\bm{r}}\right)^2\,,
\end{split}
\end{equation}

\begin{equation}
\begin{split}
    T_{\text{can}}^{0i} &= \frac{1}{2\Theta}\Biggl\{-2f_\pi^2\frac{\sin^2F}{r} + 4f_\pi^2\frac{(\xi_1 - 1 + \cos F)(1 - \cos F)}{r} - \frac{\omega^\prime\phi}{r^2} \\
    &\qquad - \frac{2G[G(1 - \xi_1) + \xi_2]}{g^2r^3} + \frac{2G\xi_1^\prime}{g^2r^2} - \frac{3g}{4\pi^2r^3}\phi F^\prime\sin^2F\Biggr\}\epsilon^{ijk}\hat{r}^jJ^k\,,
\end{split}
\end{equation}

\begin{equation}
\begin{split}
    T_{\text{can}}^{i0} &= \frac{1}{2\Theta}\Biggl\{-2f_\pi^2\frac{\sin^2F}{r} - 4f_\pi^2\frac{(G + 1 - \cos F)(1 - \cos F)}{r} - \frac{2G^2(G + 2)}{g^2r^3} \\
    &\qquad - \frac{1}{2\Theta^2}\frac{4[G(1 - \xi_1 - \xi_2) - \xi_2]\xi_2}{g^2r}(\bm{J}\cdot\hat{\bm{r}})^2 - \frac{3g}{2\pi^2r}\omega F^\prime\sin^2F\Biggr\}\epsilon^{ijk}\hat{r}^jJ^k\,.
\end{split}
\end{equation}

%%%%%
\subsection{Belinfante EMT}\label{app:formula_Tkin}

The Belinfante EMT $T^{\mu\nu}_\text{Bel}$ is obtained by adding a total-derivative term constructed from the spin current, such that the resulting tensor is symmetric.  The explicit expressions given below are arranged in the same manner as in the canonical case:

\begin{equation}
\begin{split}
    T_{\text{Bel}}^{00} &= \frac{1}{2}f_\pi^2\left({F^\prime}^2 + 2\frac{\sin^2 F}{r^2}\right) + 2f_\pi^2\frac{(G + 1 - \cos F)^2}{r^2} + f_\pi^2 g^2\omega^2 + \frac{1}{2}{\omega^\prime}^2 + \frac{{G^\prime}^2}{g^2r^2} \\
    &\qquad + \frac{G^2(G + 2)^2}{2g^2r^4} + f_\pi^2m_\pi^2(1 - \cos F) \\
    &\qquad + \frac{1}{2\Theta^2}\Biggl\{f_\pi^2\sin^2F + 2f_\pi^2(\xi_1 - 1 + \cos F)^2 + \frac{1}{2}f_\pi^2g^2\frac{\phi^2}{r^2} + \frac{\left[G(1 - \xi_1) + \xi_2\right]^2}{g^2r^2} \\
    &\qquad + \frac{1}{g^2}{\xi_1^\prime}^2 + \frac{{\phi^\prime}^2}{4r^2}\Biggr\}\left[\bm{J}^2 - (\bm{J}\cdot\hat{\bm{r}})^2\right] \\
    &\qquad + \frac{1}{2\Theta^2}\Biggl\{2f_\pi^2(\xi_1 + \xi_2)^2 + \frac{2\left[G(1 - \xi_1 - \xi_2) - \xi_2\right]^2}{g^2r^2} + \frac{1}{g^2}(\xi_1^\prime + \xi_2^\prime)^2 + \frac{\phi^2}{r^4}\Biggr\}\left(\bm{J}\cdot\hat{\bm{r}}\right)^2\,,
\end{split}
\end{equation}

\begin{equation}
\begin{split}
    T_{\text{Bel}}^{ij} &= - \frac{1}{2}f_\pi^2{F^\prime}^2\delta^{ij} + f_\pi^2\left({F^\prime}^2 - \frac{\sin^2 F}{r^2}\right)\hat{r}^i\hat{r}^j + f_\pi^2 g^2\omega^2\delta^{ij} - 2f_\pi^2\frac{(G + 1 - \cos F)^2}{r^2}\hat{r}^i\hat{r}^j \\
    &\qquad + \frac{1}{2}{\omega^\prime}^2(\delta^{ij} - 2\hat{r}^i\hat{r}^j) + \frac{{G^\prime}^2}{g^2r^2}\hat{r}^i\hat{r}^j + \frac{G^2(G + 2)^2}{2g^2r^4}(\delta^{ij} - 2\hat{r}^i\hat{r}^j) \\
    &\qquad - f_\pi^2m_\pi^2(1 - \cos F)\delta^{ij} \\
    &\qquad + \frac{1}{2\Theta^2}\Biggl\{f_\pi^2\sin^2F\delta^{ij} + 2f_\pi^2(\xi_1 - 1 + \cos F)^2\delta^{ij} + \frac{1}{2}f_\pi^2g^2\frac{\phi^2}{r^2}(\delta^{ij} - 2\hat{r}^i\hat{r}^j) \\
    &\qquad + \frac{\left[G(1 - \xi_1) + \xi_2\right]^2}{g^2r^2}\delta^{ij} + \frac{1}{g^2}{\xi_1^\prime}^2(\delta^{ij} - 2\hat{r}^i\hat{r}^j) + \frac{{\phi^\prime}^2}{4r^2}\delta^{ij}\Biggr\}\left[\bm{J}^2 - (\bm{J}\cdot\hat{\bm{r}})^2\right] \\
    &\qquad + \frac{1}{2\Theta^2}\Biggl\{2f_\pi^2(\xi_1 + \xi_2)^2\delta^{ij} - f_\pi^2g^2\frac{\phi^2}{r^2}(\delta^{ij} - \hat{r}^i\hat{r}^j) + \frac{1}{g^2}(\xi_1^\prime + \xi_2^\prime)^2(\delta^{ij} - 2\hat{r}^i\hat{r}^j) \\
    &\qquad - \frac{2[G(1 - \xi_1) + \xi_2]^2}{g^2r^2}(\delta^{ij} - 2\hat{r}^i\hat{r}^j) - \frac{2G(G + 2)(1 - \xi_1)\xi_2}{g^2r^2}\hat{r}^i\hat{r}^j \\
    &\qquad - \frac{2G(G + 2)(1 - \xi_1 - \xi_2)\xi_2}{g^2r^2}\hat{r}^i\hat{r}^j + \frac{\phi^2}{r^4}\delta^{ij} - \frac{{\phi^\prime}^2}{2r^2}\delta^{ij} - \frac{1}{2}\left(\frac{\phi^\prime}{r} - \frac{2\phi}{r^2}\right)^2\hat{r}^i\hat{r}^j \\
    &\qquad + \frac{\phi^\prime}{r}\left(\frac{\phi^\prime}{r} - \frac{2\phi}{r^2}\right)\hat{r}^i\hat{r}^j\Biggr\}\left(\bm{J}\cdot\hat{\bm{r}}\right)^2\,,
\end{split}
\end{equation}

\begin{equation}
\begin{split}
    T_{\text{Bel}}^{0i} = T_{\text{Bel}}^{i0} &= \frac{1}{2\Theta}\Biggl\{-2f_\pi^2\frac{\sin^2F}{r} + 4f_\pi^2\frac{(G + 1 - \cos F)(\xi_1 - 1 + \cos F)}{r} - 2f_\pi^2g^2\frac{\omega\phi}{r} \\
    &\qquad - \frac{2G(G + 2)[G(1 - \xi_1) + \xi_2]}{g^2r^3} - \frac{\omega^\prime\phi^\prime}{r} + \frac{2G^\prime\xi_1^\prime}{g^2r}\Biggr\}\epsilon^{ijk}\hat{r}^jJ^k\,.
\end{split}
\end{equation}

%%%%%%%%%%
\section{Analytical expressions of the angular-momentum tensor in the model}
\label{app:formula_Scan}

We also provide analytical expressions for the canonical spin current tensor $S^{\lambda\mu\nu}$ in our model.  By definition, $S^{\lambda\mu\nu}$ is antisymmetric in its last two indices, $S^{\lambda\mu\nu}=-S^{\lambda\nu\mu}$, and enters the canonical decomposition of the total angular-momentum current.  Here, we show $S^{\lambda\mu\nu}$ only, because the OAM, $L^{\lambda\mu\nu}$, can be readily constructed from $T^{0i}$ given in App.~\ref{app:formula_Tcan}, and the total angular momentum $J^{\lambda\mu\nu}$ is then obtained by adding the spin contribution $S^{\lambda\mu\nu}$ to $L^{\lambda\mu\nu}$.

The following expressions summarize the nonvanishing components of the canonical spin current tensor:

\begin{equation}
\begin{split}
    S^{0ij} &= \frac{1}{2\Theta}\Biggl\{\frac{4G[G(1 - \xi_1 - \xi_2) - \xi_2]}{g^2r^2}(\bm{J}\cdot\bm{\hat{r}})\varepsilon^{ijl}\hat{r}^l - \frac{2G\xi_1^\prime}{g^2r}(\hat{r}^i\varepsilon^{jlk} - \hat{r}^j\varepsilon^{ilk})\hat{r}^lJ^k \\ 
    &\qquad + \frac{\omega^\prime\phi}{r}(\hat{r}^i\varepsilon^{jlk} - \hat{r}^j\varepsilon^{ilk})\hat{r}^lJ^k\Biggr\}\,,
\end{split}
\end{equation}

\begin{equation}
\begin{split}
    S^{00i} &= - \omega\omega^\prime\hat{r}^i - \frac{1}{2\Theta^2}\Biggl\{\frac{2}{g^2}\xi_1\xi_1^\prime\hat{r}^i[\bm{J}^2 - (\bm{J}\cdot\hat{\bm{r}})^2] + \frac{2}{g^2}(\xi_1 + \xi_2)(\xi_1^\prime + \xi_2^\prime)\hat{r}^i(\bm{J}\cdot\hat{\bm{r}})^2\Biggr\}\,,
\end{split}
\end{equation}

\begin{equation}
\begin{split}
    S^{kij} &= \frac{GG^\prime}{g^2r^2}(\hat{r}^i\delta^{jk} - \delta^{ik}\hat{r}^j) + \frac{1}{2\Theta^2}\Biggl\{-\frac{\phi}{2r^2}\left(\frac{2\phi}{r} - \phi^\prime\right)(\hat{r}^i\delta^{jk} - \delta^{ik}\hat{r}^j)[\bm{J}^2 - (\bm{J}\cdot\hat{\bm{r}})^2] \\
    &\qquad + \frac{\phi^2}{r^3}(\hat{r}^i\delta^{jk} - \delta^{ik}\hat{r}^j)[\bm{J}^2 - (\bm{J}\cdot\hat{\bm{r}})^2] - \frac{\phi\phi^\prime}{2r^2}(\hat{r}^i\delta^{jk} - \delta^{ik}\hat{r}^j)(\bm{J}\cdot\hat{\bm{r}})^2\Biggr\}\,,
\end{split}
\end{equation}

\begin{equation}
\begin{split}
    S^{k0i} &= \frac{1}{2\Theta}\Biggl\{\frac{2\xi_1}{g^2r}\left(\frac{2G}{r} - G^\prime\right)(\varepsilon^{ijl}\hat{r}^k - \varepsilon^{kjl}\hat{r}^i)\hat{r}^jJ^l + \frac{4G\xi_1}{g^2r^2}\varepsilon^{kij}J^j + \frac{2G^2\xi_1}{g^2r^2}\varepsilon^{kij}\hat{r}^l\hat{r}^jJ^l \\
    &\qquad + \frac{2G(G + 2)\xi_2}{g^2r^2}\varepsilon^{kij}\hat{r}^j\hat{r}^lJ^l - \frac{2G[G(1 - \xi_1 - \xi_2) - \xi_2]}{g^2r^2}\varepsilon^{kij}\hat{r}^j\hat{r}^lJ^l \\
    &\qquad - \frac{\omega}{r}\left(\frac{2\phi}{r} - \phi^\prime\right)(\varepsilon^{ijl}\hat{r}^k - \varepsilon^{kjl}\hat{r}^i)\hat{r}^jJ^l - \frac{2\omega\phi}{r^2}\varepsilon^{kij}J^j \\
    &\qquad + \frac{2G\xi_1^\prime}{g^2r}\hat{r}^k\varepsilon^{ijl}\hat{r}^jJ^l - \frac{\omega^\prime\phi}{r}\hat{r}^k\varepsilon^{ijl}\hat{r}^jJ^l\Biggr\}\,.
\end{split}
\end{equation}

\bibliography{Skyrme}

\end{document}